\documentclass[reprint,nofootinbib,pre]{revtex4-1}
\usepackage{hyperref}
\usepackage{subfig}
\usepackage{graphicx}
\usepackage{amsmath,amssymb}
\usepackage{algorithm,algorithmic}
\usepackage{units}

\begin{document}
\title{Thermodynamic phase-field model for microstructure with multiple components and phases: the possibility of metastable phases}
\author{Daniel A. Cogswell}
\author{W. Craig Carter}
\affiliation{Department of Materials Science \& Engineering, Massachusetts Institute of Technology, Cambridge, MA 02139, USA}

\date{\today}

\begin{abstract}
A diffuse-interface model for microstructure with an arbitrary number of components and phases was developed from basic thermodynamic and kinetic principles and formalized within a variational framework.  The model includes a composition gradient energy to capture solute trapping, and is therefore suited for studying phenomena where the width of the interface plays an important role.  Derivation of the inhomogeneous free energy functional from a Taylor expansion of homogeneous free energy reveals how the interfacial properties of each component and phase may be specified under a mass constraint.  A diffusion potential for components was defined away from the dilute solution limit, and a multi-obstacle barrier function was used to constrain phase fractions.  The model was used to simulate solidification via nucleation, premelting at phase boundaries and triple junctions, the intrinsic instability of small particles, and solutal melting resulting from differing diffusivities in solid and liquid.  The shape of metastable free energy surfaces is found to play an important role in microstructure evolution and may explain why some systems premelt at phase boundaries and phase triple junctions while others do not.
\end{abstract}

\maketitle

\section{Introduction}
Developing an understanding of microstructure formation in multiphase, multicomponent systems is a challenge for industrial development of advanced alloys, yet interesting from a philosophical perspective due to the formation of complex patterns for which no theory exists \cite{Hecht2004}.  Compared to two-phase binary systems, multiphase and multicomponent systems have additional degrees of freedom that introduce inherent complexity.  Multiphase systems have the ability to form phase triple junctions, transient phases \cite{Macdonald1992}, and metastable phases at grain boundaries or triple junctions (i.e. interfacial premelting) \cite{Hsieh1989,Mei2007}.

\begin{figure}[h]
 \includegraphics[width=.65\columnwidth]{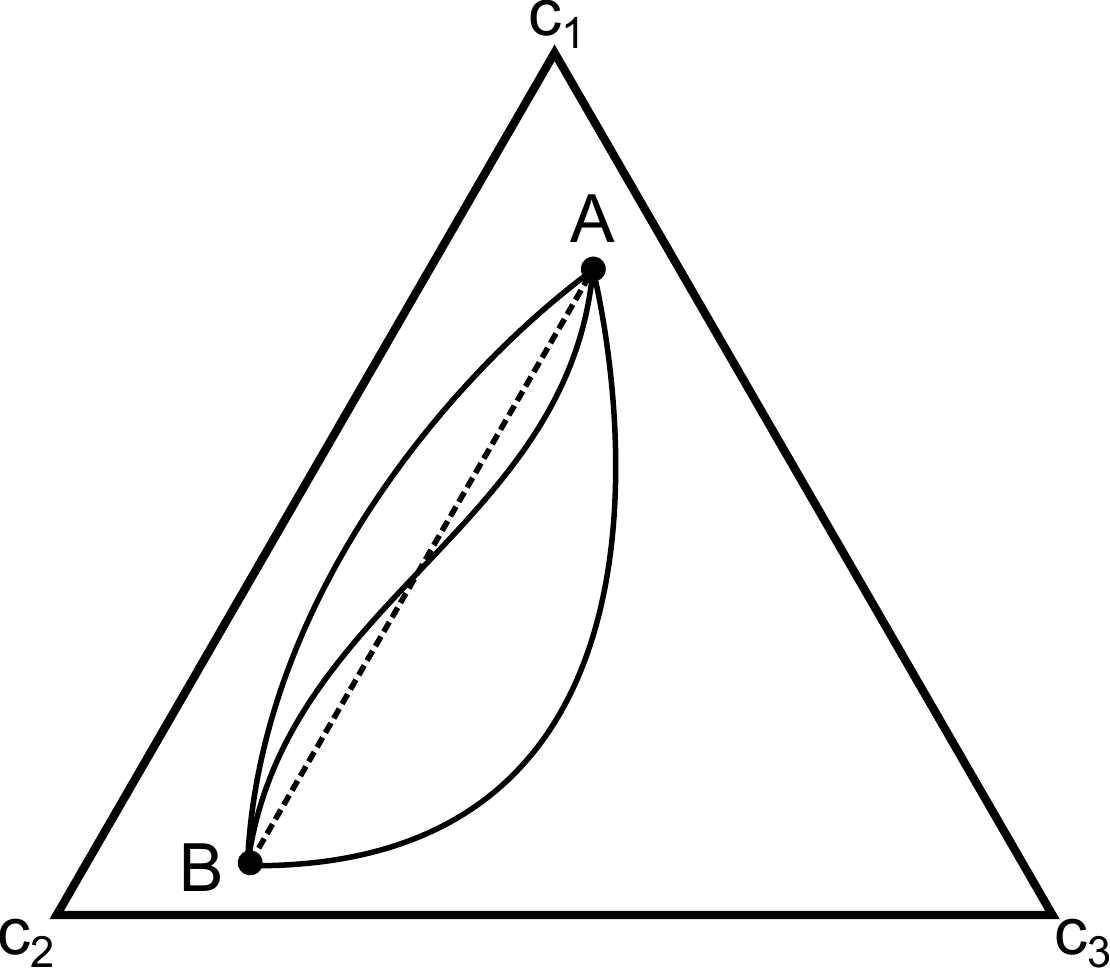}
 \caption{A hypothetical ternary phase diagram.  A and B denote the bulk composition of two phases in equilibrium.  Diffuse interfaces are curves that connect A and B.}
 \label{Fig:motivation}
\end{figure}

Additionally in multicomponent systems, adsorption of components to interfaces or triple junctions is possible.  Consider figure \ref{Fig:motivation}, which illustrates a hypothetical ternary phase diagram.  A and B denote the bulk composition of two phases in equilibrium, and a diffuse interface between A and B corresponds to a curve connecting the points.  At equilibrium there can be only one curve.  In binary systems, the interfacial profile is constrained to lie on the dotted line that represents a linear combination of the components.  In a ternary system however, the additional component permits complex pathways that depend on the free energies of the phases, the presence of metastable phases, and the energy-minimizing path through composition space.

In this work we develop a multicomponent, multiphase model that treats the diffuse interface in a thermodynamically consistent way, allowing us to investigate premelting and the effects of metastable phases on interfacial composition in multiphase systems.  The model includes a $(\nabla c)^2$ in the free energy functional as a natural way to model the correct amount of solute trapping \cite{Wheeler1993}.  We present a careful derivation of the model in order to cast the multiphase, multicomponent problem within a thermodynamic framework, and derive component diffusion equations that obey the Gibbs-Duhem and Nernst-Einstein relations.  Subtle differences are clarified between the chemical potential and the diffusional potential.  These differences become important in multicomponent systems.  Specifically, the driving force for diffusion is sometimes defined as $\frac{\delta F}{\delta c_i}$  \cite{Eyre1993,Elliott1997,Cha2001,Zhang2006}, but this definition is incorrect for a multicomponent system that obeys a mole fraction constraint.

Phase-field  has emerged as an important method for modeling microstructure evolution because of its ability to simulate complex geometries while incorporating thermodynamic and kinetic data.  A phase-field model assumes that interfaces in microstructure are diffuse at the nanoscale and can be represented by one or more smoothly varying order parameters, eliminating the need to explicitly track boundaries.  Nonlinear diffusion and curvature-driven physics are incorporated, and creation, destruction, and merging of interfaces are handled implicitly.  A substantial amount of literature has been written on phase-field models and is summarized in recent review papers \cite{Boettinger2002,Chen2002,Moelans2008,Singer-Loginova2008,Steinbach2009}.

Phase-field models may be classified into two categories based on their philosophical treatment of a diffuse interface.  In the approach pioneered by Cahn and Hilliard \cite{Cahn1958}, the interface is a coarse-graining of the underlying atomistic representation.  The width of the interface in the model is identical to its physical width, which may be as small as a few nanometers.  This approach produces a thermodynamically consistent description of the interface, but makes simulation of realistically sized microstructures problematic.  Microstructural features are often on the order of micrometers or larger, several orders of magnitude larger than the interfacial width.  This presents a computational challenge for simulating large microstructure.  In principle, the problem could be addressed with faster computers and improved numerical algorithms.

The second approach uses phase-field as a modeling tool for solving the underlying free boundary problem without explicitly tracking boundaries.  It is numerically advantageous to allow the computational interfacial width $W$ to exceed the physical width, but doing so introduces error that scales with $W$  \cite{Folch2005}.  The ``thin interface'' approach \cite{Karma1996,Karma1998} was an important development in this regard.  With the appropriate choice of model parameters, thin interface models converge to the Gibbs-Thomson relation in the limit where the interfacial width is much smaller than a typical pattern size of the system.  As a result, convergence is on the order of $W^2$.  Excessive solute trapping occurs when the numerical width becomes large, but has been remedied with anti-trapping currents \cite{Karma2001}.  Notably, this approach has produced simulations of dendrites that are quantitatively comparable to experiment.

Alloy phase-field models have been developed following both philosophies and will be briefly reviewed.   Wheeler, Boettinger, and McFadden (WBM) \cite{Wheeler1992,Wheeler1993,Wheeler1996} treated the interface in a thermodynamic way but were limited to binary systems with two free energy curves due to fundamental model difficulties.   Steinbach et al. \cite{Steinbach1996,Steinbach1999} prompted development of a series of models for multicomponent and multiphase systems that  have produced quantitative simulations on experimental length scales \cite{Asta2009}.  However these multiphase models are not appropriate for studying phenomena where the interfacial width plays an important role, such as solute trapping, interface premelting, nucleation, or the appearance of transient phases.

\subsection{The Wheeler-Boettinger-McFadden model}
The Wheeler-Boettinger-McFadden (WBM) model \cite{Wheeler1992,Wheeler1993,Wheeler1996} begins with the Cahn-Hilliard free energy functional for a binary system \cite{Cahn1958} (see section \ref{Sec:Binary_free_energy}) and introduces a non-conserved order parameter $\phi$ to indicate which regions of the system are solid ($\phi=1$) and which are liquid ($\phi=0$).  At an interface between liquid and solid, both $\phi$ and $c$ vary smoothly from one phase to the other.  The free energy functional for the system is:
\begin{equation}
 F[c,\phi]=\int_V\left[f_0(\phi,c,T)+\frac{1}{2}\epsilon_c(\nabla c)^2+\frac{1}{2}\epsilon_\phi(\nabla \phi)^2\right]dV
\end{equation}
The homogeneous free energy density $f_0(\phi,c,T)$ promotes phase separation in the absence of interfacial energies, and $\epsilon_c$ and $\epsilon_\phi$ are the composition and phase gradient energies, respectively.  Phase and composition gradients overlap at equilibrium to form an interface, and the gradient squared terms smooth the interface and introduce interfacial energy.  The $(\nabla c)^2$ term was omitted from the original model for computational convenience \cite{Wheeler1992}, but was later found to be necessary for modeling solute drag during rapid solidification \cite{Wheeler1993}.

The WBM approach models a diffuse interface as an interpolation between phases where the composition of phases at the interface are equal.  An interpolating function $p(\phi)$ is used to connect the homogeneous free energy densities of the phases:
\begin{equation}
 f_0(\phi,c,T)=p(\phi)f_{liq}(c,T)+(1-p(\phi))f_{sol}(c,T)
\end{equation}
Interpolation between two free energy curves is illustrated in figure \ref{Fig:interpolation}.  $p(\phi)$ has a minimum at $\phi=0$ and $\phi=1$ and provides a barrier for transition from one phase to the other.  It lacks a physical basis and is generally chosen for numerical convenience.  The WBM model requires that $p(\phi)$ approach $\phi=0$ and $\phi=1$ with zero slope in order to prevent the appearance of negative phase fractions.

Because there is no natural extension of $p(\phi)$ to handle an arbitrary number of free energy curves, there have been few attempts to develop a multiphase model following the WBM approach.  Folch and Plapp \cite{Folch2005} derived a thin-interface model that included an interpolating function for three curves, and Nestler et al. \cite{Nestler2005} developed a WBM-like nonisothermal multiphase, multicomponent model.  However, neither work included a composition gradient energy or applied the correct thermodynamic constraints to the component diffusion equations.  These issues are addressed in sections \ref{Sec:Free_energy_functional} and \ref{Sec:Component_evolution}.

\subsection{The Access multiphase model}
Steinbach and co-workers developed the so-called Access multiphase model, the first phase-field model capable of simulating the interaction of an arbitrary number of phases \cite{Steinbach1996,Steinbach1999}.  The original model did not include solute diffusion and considered pairwise interactions between phases using double well interpolation functions and Allen-Cahn dynamics.  Modeling the dynamics of a multiphase system as the sum of pair-wise interactions was problematic, violating Young's Law at phase triple junctions, and was fixed with the introduction of interface fields \cite{Steinbach1999}.

Tiaden et al. \cite{Tiaden1998} added single-component solute diffusion to the Access multiphase model.  The interfacial region was modeled as a blend of phases, each with a phase fraction $\phi_\alpha$ and unique composition $c_\alpha$, constrained so that the concentration of the system was $c(x,t)=\sum_\alpha\phi_\alpha c_\alpha$.  The diffusing species was partitioned amongst the different phases, and Fickian diffusion equations were solved in each phase.  The diffusion equations were coupled to phase evolution equations driven by a difference in free energy between phases which was determined from a local linearization of the phase diagram.

The dilute solution limitation of the Tiaden model was removed in an extension by Kim et al. \cite{Kim1999} for single-component diffusion with the use of an interpolating function to link the free energy curves.  Kim also introduced a more sophisticated condition of equal chemical potential to determine how to distribute solute amongst the phases at a diffuse interface.

Grafe et al. \cite{Grafe2000} developed the first multicomponent extension of the Tiaden model.  The driving force for diffusion was $\nabla c_i^\alpha$, the concentration gradient of component $i$ in phase $\alpha$, which is a dilute solution approximation.  Solute distribution was calculated with partition coefficients from Thermo-Calc.

Eiken et al. \cite{Eiken2006} developed a multicomponent extension of the Tiaden model which removed the dilute solution limitation and allowed for easier inclusion of thermodynamic data.  $\widetilde{\mu}_i^\alpha=\frac{\partial \overline{G}_\alpha}{\partial c_i^\alpha}$ was chosen as the driving force for diffusion, although $\widetilde{\mu}_i^\alpha$ is the slope of the free energy curves and not the chemical potential $\mu_i^\alpha=\frac{\partial G_\alpha}{\partial n_i^\alpha}$ (see section \ref{Sec:generalized_diffusion_potential}).  A very computationally expensive quasi-equilibrium calculation was necessary at each timestep to relieve the dilute solution approximation.

A two-phase multicomponent model with an antitrapping current was presented by Kim \cite{Kim2007}, but to date an antitrapping current has not been included in a model with both an arbitrary number of phases and components.

\subsection{Graphical interpretation}

\begin{figure}[t]
 \centerline{
 \subfloat[Molar free energy for two phases, $\alpha$ and $\beta$.]{
  \includegraphics[width=.75\columnwidth]{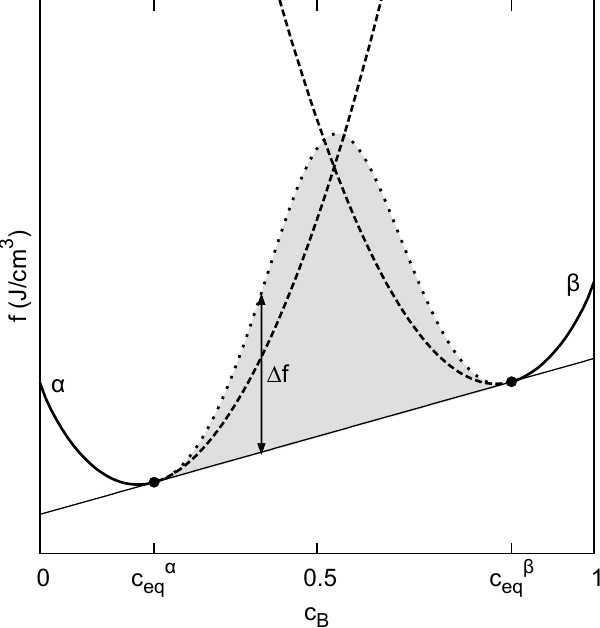}
  \label{Fig:delta_G}}}
 \centerline{
 \subfloat[An interpolating function is used to smoothly connect the molar free energy curves.]{
  \includegraphics[width=.8\columnwidth]{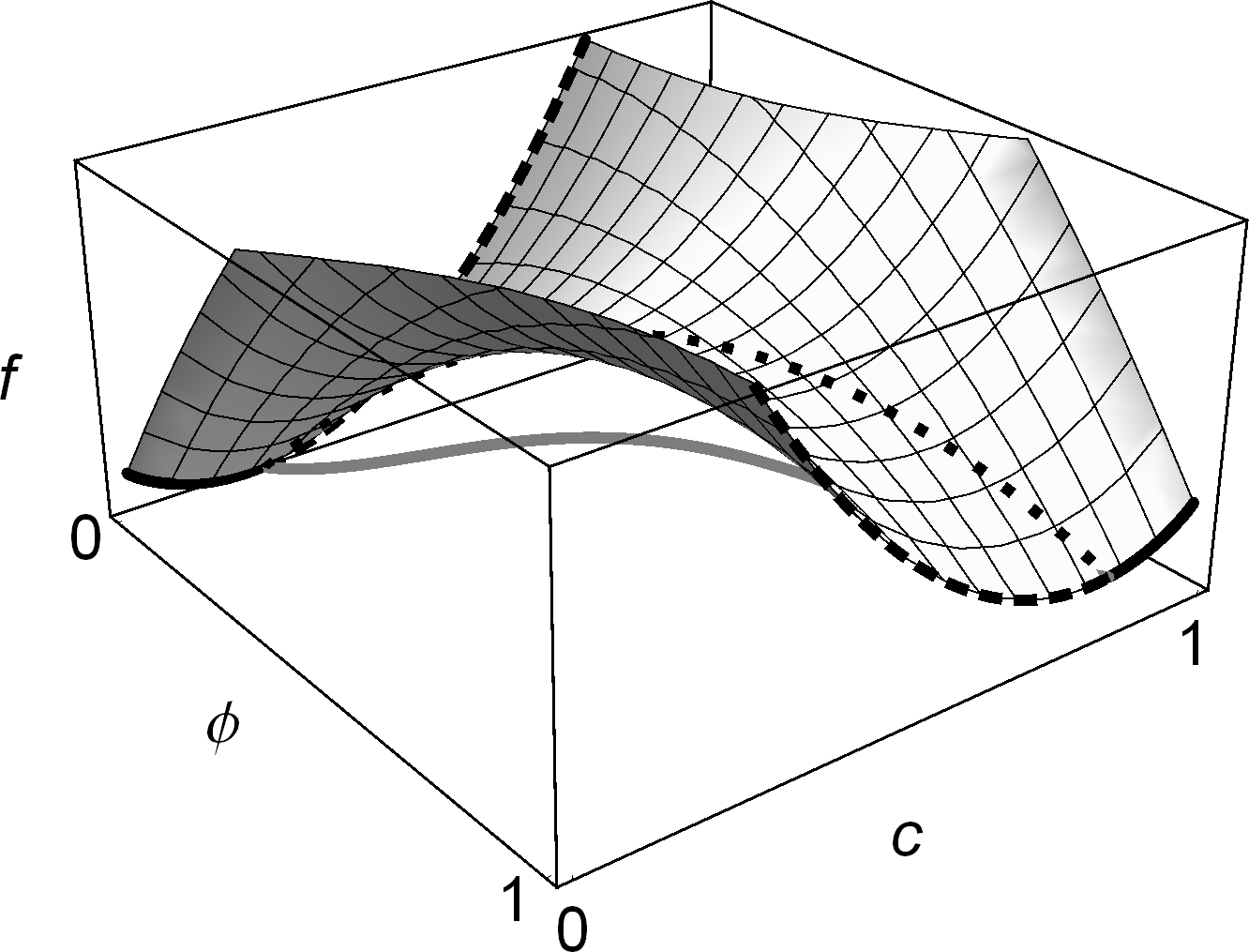}
  \label{Fig:interpolation}}}
 \caption{The WBM model assumes that the energy of interfacial compositions is a weighting of the dashed regions of the free energy curves, while Access models assume that the energies of interfacial compositions lie on the common tangent line.}
\end{figure}

The fundamental difference between the WBM and Access approaches is illustrated in figure \ref{Fig:delta_G}, where free energy curves and the common tangent construction for two phases $\alpha$ and $\beta$ are drawn.  A diffuse interface must include compositions between the equilibrium concentrations $c_{eq}^{\alpha}$ and $c_{eq}^{\beta}$, but the energy of these intermediate compositions is somewhat ambiguous.

The WBM model assumes that each phase at an interface has the same composition.   At equilibrium, the free energy of these interfacial points is then a weighted average of the dashed portions of the free energy curves in figure \ref{Fig:delta_G}.  The dotted line in figure \ref{Fig:delta_G} indicates a potential path when a barrier in $\phi$ is added, and corresponds to the dotted path lying on the free energy surface in figure \ref{Fig:interpolation}.  When the system is not at equilibrium, interfacial compositions may lie anywhere on the surface of figure \ref{Fig:interpolation}.  $\Delta f$ denotes energy at an interfacial point relative to a composite blend of $\alpha$ and $\beta$ at equilibrium.  The gray shaded region is $\Delta f$ integrated across an interface, and represents the interfacial energy contribution from including intermediate compositions at the interface.  The contribution of the shaded area increases for wider interfaces because more material with energy above the common tangent construction must be introduced.

The Access approach for modeling interfaces is to assume each phase has its own unique composition that cannot be measured experimentally but which evolves toward its equilibrium concentration.  Interpolation between phases at their equilibrium concentration produces intermediate compositions with energies that lie on the common tangent line with a barrier only in $\phi$, prohibiting the appearance of metastable phases which lie above the common tangent.

The gray line in figure \ref{Fig:interpolation} illustrates an Access interface that connects the common tangent points of the free energy curves.  Because the barrier in $\phi$ is the only contributor to $\Delta f$, widening the interface for computational convenience does not introduce more material with energy above the common tangent.

\section{The multiphase free energy functional}
\label{Sec:Free_energy_functional}
\subsection{Free energy of a binary system}
\label{Sec:Binary_free_energy}
In an influential paper that laid the foundation for phase-field modeling, Cahn and Hilliard derived an expression for the free energy of an inhomogeneous binary system \cite{Cahn1958}.  Their approach was to assume that the free energy of an infinitesimal volume in a nonuniform system depends both on its composition and the composition of its nearby environment.  Total free energy cannot depend solely on local composition because different spatial configurations with the same volume fraction are not energetically equivalent; a heterogeneous system has more interfacial area and will have a higher energy.

Starting with the homogeneous free energy density for a binary system $f_0(c)$, they performed a Taylor expansion in terms of the derivatives of composition to approximate $f(c,\nabla c,\nabla^2 c,\ldots)$.   Morris \cite{Morris1971} provides justification for excluding terms linear in $\vert\nabla c\vert$.  For isotropic or cubic symmetry  $(\partial f/\partial\nabla c)_0=0$, and the free energy simplifies to an equation with constant coefficients and even powers of $\nabla c$:
\begin{equation}
  f=f_0(c)+\kappa_1\nabla^2 c+\frac{1}{2}\kappa_2(\nabla c)^2+\frac{1}{2}\kappa_3(\nabla^2 c)^2+\kappa_4\nabla^4 c+\ldots
\end{equation}
Cahn and Hilliard then argued that the derivative terms with even powers $\nabla^2 c$, $\nabla^4 c$, $\nabla^6 c$, etc. should vanish.  Because of the assumption that the free energy density is influenced only by concentration within a small neighborhood, it is reasonable to truncate the expansion.  Keeping only terms up to second-order produces the Cahn-Hilliard free energy functional:
\begin{equation}
 F[c]=\int_Vf_0(c)+\kappa(\nabla c)^2\, dV
\end{equation}
where $\kappa$ is a gradient energy coefficient that penalizes the formation of sharp interfaces.

\subsection{Free energy of a multicomponent system}
\label{Sec:Multicomponent_free_energy}
The approach of Cahn and Hilliard is now applied to a system with an arbitrary number of components.  A system with $M$ components has $M-1$ independent mole fractions that obey the following constraint:
\begin{equation}
 \sum_{i=1}^M c_i=1
 \label{Eq:mass_constraint}
\end{equation}
The inhomogeneous free energy becomes a function of each independent component as well as their derivatives: $f(c_1,c_2,\ldots,\nabla c_1,\nabla c_2,\ldots,\nabla^2 c_1,\nabla^2 c_2,\ldots)$.  The Taylor expansion \cite{Corwin1995} of the multicomponent $f$ about a homogeneous point $f_0=f(c_1,c_2,\ldots,0,0,\ldots,0,0,\ldots)$ is:
\begin{equation}
 \begin{split}
  f(&c_1,c_2,\ldots,\nabla c_1,\nabla c_2,\ldots,\nabla^2 c_1,\nabla^2 c_2,\ldots)\\
  &=f_0(c_1,c_2,\ldots)\\
  &+\left(\frac{\partial f}{\partial\nabla c_1}\right)_0\nabla c_1
   +\left(\frac{\partial f}{\partial\nabla c_2}\right)_0\nabla c_2+\ldots\\
  &+\left(\frac{\partial f}{\partial\nabla^2 c_1}\right)_0\nabla^2 c_1
   +\left(\frac{\partial f}{\partial\nabla^2 c_2}\right)_0\nabla^2 c_2+\ldots\\
  &+\frac{1}{2}\left(\frac{\partial^2 f}{\partial(\nabla c_1)^2}\right)_0(\nabla c_1)^2
   +\frac{1}{2}\left(\frac{\partial^2 f}{\partial(\nabla c_2)^2}\right)_0(\nabla c_2)^2+\ldots\\
  &+\left(\frac{\partial^2 f}{\partial\nabla c_1\partial\nabla c_2}\right)_0\nabla c_1\cdot\nabla c_2+\ldots\\
  &+\ldots
 \end{split}
 \label{Eq:Multicomponent_taylor_expansion}
\end{equation}
For simplicity only terms for two components up to second-order have been written out because higher order terms will be excluded.  Once again only isotropic and cubic symmetry of $f$ is considered, allowing the tensors to be replaced with constants, and terms in $\nabla c$, as well as even derivatives, are excluded.  The Taylor expansion of f in terms of the $M-1$ independent components and their derivatives is:
\begin{equation}
  f=f_0(c_1,c_2,\ldots)+\sum_{i=1}^{M-1}\frac{1}{2}\kappa_i(\nabla c_i)^2+\sum_{j<i}\sum_{i=1}^{M-1}\kappa_{ij}\nabla c_i\cdot\nabla c_j
 \label{Eq:Multicomponent_expansion}
\end{equation}
This equation, which has been previously reported in literature \cite{ARIYAPADI1992,Eyre1993,Elliott1997}, may be simplified by combining the summation terms:
\begin{equation}
 f=f_0(c_1,c_2,\ldots)+\sum_{i,j=1}^{M-1}\frac{1}{2}\kappa_{ij}\nabla c_i\cdot\nabla c_j
 \label{Eq:Multicomponent_simplified_expansion}
\end{equation}
where $\kappa_{ij}$ is a symmetric matrix of gradient energy coefficients discussed in section \ref{Sec:gradient_energy_matrix}.  The free energy functional for a multicomponent system is:
\begin{equation}
 F[\{c\}]=\int_V \bigg[f_0(\{c\})+\sum_{i,j=1}^{M-1}\frac{1}{2}\kappa_{ij}\nabla c_i\cdot\nabla c_j)\bigg]\, dV
 \label{Eq:F_multicomponent}
\end{equation}
where $\{c\}$ denotes a set of $M-1$ independent mole fractions.

\subsection{Definition of a phase}
A phase is a region of a microstructure with homogeneous properties that is physically distinct from other regions of the system, excluding geometric transformations that map one region onto another.  Phases in microstructure commonly differ in composition and/or crystal structure, although many other physical differences are possible.  The volume fraction of phases in equilibrium is predicted from thermodynamics, but phase itself is not a thermodynamic state variable; phase is a labeling device that identifies a unique thermodynamic state function.

Each phase $\alpha$ is assigned a phase fraction $\phi_\alpha$ that varies between 0 and 1.  $\phi_\alpha=0$ designates areas where no $\alpha$-phase is present, and $\phi_\alpha=1$ corresponds to single-phase regions of $\alpha$.  For a system with $N$ phases, the phase fractions obey the following constraint:
\begin{equation}
 \sum_{\alpha=1}^N\phi_\alpha=1
 \label{Eq:phase_constraint} 
\end{equation}

Microstructure (excluding grain boundaries, defects, etc.) is composed of single phase regions separated by interfaces, and only at interfaces are more than one $\phi$ non-zero.  The interface between two phases is assumed to consist of a thin layer across which the physical properties vary continuously from one phase to the other, and a diffuse interface at equilibrium represents a balance between free energy curves, composition gradients, and phase gradients.

Because the thermodynamic potential of a multiphase system is equal to the summation of potentials over all phases, a linear weighting of the free energy densities by phase fractions is used to represent the homogeneous free energy of a multiphase system:
\begin{equation}
 f_0(\{c\},\{\phi_1,\phi_2,\ldots\phi_N\})=\sum_{\alpha=1}^N\phi_{\alpha}f_{\alpha}(\{c\})
 \label{Eq:linear_weighting}
\end{equation}
This form reduces to $f_\alpha(\{c\})$ when only the $\alpha$-phase is present, yet can be constructed for an arbitrary number of phases.

Eq. \ref{Eq:linear_weighting} does not provide an energy barrier for diffusionless phase transformations, and without modification does not correctly describe phase transitions of pure components.  A simple barrier between $\alpha$ and $\beta$ of the form $W_{\alpha\beta}\phi_\alpha\phi_\beta$ is suggested, although the multi-obstacle barrier introduced in section \ref{Sec:phase_barrier_function} permits \textit{any} function to be used as a barrier.  The homogeneous free energy becomes:
\begin{equation}
 f_0(\{c\},\{\phi\},T)=\sum_{\alpha=1}^N\phi_{\alpha}f_{\alpha}(\{c\},T)+\sum_{\beta\ne\alpha}W_{\alpha\beta}\phi_\alpha\phi_\beta
 \label{Eq:f_barrier}
\end{equation}
$W_{\alpha\beta}>0$ captures the mean field interaction between phases and is analogous to a positive enthalpy of mixing for phases.

\subsection{Free energy of a multiphase, multicomponent system}
\label{Sec:multiphase_multicomponent}
For an $N$-phase, $M$-component system, $f(\{c\},\{\phi\},\{\nabla c\},\{\nabla\phi\},\ldots)$ is a function of $M-1$ independent mole fractions and $N-1$ independent phase fractions.  Because $f$ can describe nonequilibrium systems where metastable phases are present, it is not necessary that the Gibbs phase rule be obeyed.

Once again isotropic and cubic symmetry of the free energy is considered, terms in $\nabla c$ and even derivatives of $c$ are excluded from the Taylor expansion, and only terms up to second-order are kept.  The full expansion about the homogeneous free energy $f_0(\{c\},\{\phi\},\{0\},\{0\},\ldots)$ is not algebraically difficult but has many terms and is not explicitly written out here.  It is analogous to Eq. \ref{Eq:F_multicomponent} but with two additional sets of terms.  One set couples phase gradients with gradient energy coefficients $\lambda_{\alpha\beta}$.  The second set couples composition gradients and phase gradients:
\begin{equation}
 \left(\frac{\partial^2f}{\partial\nabla c_i\partial\nabla \phi_\alpha}\right)_0\nabla c_i\cdot\nabla\phi_\alpha
\end{equation}
The coefficients of these terms form an $M\times N$ matrix $\xi_{i\alpha}$, and introduce an additional energy penalty for overlapping phase and concentration gradients.  The total gradient energy contribution for a multicomponent, multiphase system may be written in compact form as:
\begin{equation}
 \frac{1}{2}
 \begin{bmatrix}\nabla c & \nabla\phi\end{bmatrix}
 \begin{bmatrix}
  \kappa_{ij} & \xi_{i\alpha}\\
  \xi_{\alpha i} & \lambda_{\alpha\beta}
 \end{bmatrix}
 \begin{bmatrix}\nabla c\\ \nabla \phi\end{bmatrix}
 \label{Eq:gradient_energy}
\end{equation}

For simplicity we assume that $\xi_{i\alpha}=0$ in this work.  The multiphase, multicomponent free energy functional then becomes:
\begin{equation}
\begin{split}
 F[\{c\},\{\phi\}]=\int_V \bigg[f_0&+\sum_{\alpha,\beta=1}^{N-1}\frac{1}{2}\lambda_{\alpha\beta}\nabla\phi_\alpha\cdot\nabla\phi_\beta\\
 &+\sum_{i,j=1}^{M-1}\frac{1}{2}\kappa_{ij}\nabla c_i\cdot\nabla c_j\bigg]\, dV
 \label{Eq:F_multiphase}
\end{split}
\end{equation}
The free energy curves are the driving force for phase separation, and the gradient energy coefficients $\lambda_{\alpha\beta}$ and $\kappa_{ij}$ penalize gradients that develop, creating a surface energy at phase boundaries.  $\kappa_{ij}$ penalizes phases for differing in composition, and $\lambda_{\alpha\beta}$ introduces additional energy not captured by the composition gradients at phase boundaries.  This energy derives from some physical difference between the phases other than composition.  Eq. \ref{Eq:F_multiphase}, which is the central equation of focus in this work, is a first order approximation of the free energy of a system with an inhomogeneous distribution of phases and components.  It reduces to the Cahn-Hilliard equation for a two-component system.

\subsection{Surface energy}
The surface energy of a diffuse interface is the excess grand canonical potential, the difference between the free energy functional and the minimized free energy the system would have if the properties of the phases were continuous:
\begin{equation}
 \sigma=\min F[\{c\},\{\phi\}]-\sum_i\mu_i^en_i
 \label{Eq:interfacial_energy}
\end{equation}
The first term is the minimum of $F$ found by application of the Euler-Lagrange equation, and $\sum_i\mu_i^ec_i$ is the homogeneous free energy.  $\mu_i^e$ is the chemical potential of component $i$ at equilibrium and is found by computing the tangent plane to the free energy surfaces.

Surface energy in this model has two contributions.  One contribution comes from phase and composition gradients which are present at the interface, and the other results from composition deviating from its equilibrium value at the interface as illustrated in figure \ref{Fig:delta_G}.

\subsection{Interpretation of the gradient energy matrices}
\label{Sec:gradient_energy_matrix}
Although Eq. \ref{Eq:F_multiphase} is a function only of the independent gradients, it is necessary to specify the properties of the dependent component and phase as well.\footnote{We found that ignoring the dependent phase and component produced an unexpected asymmetry in interfacial compositions that was visible in composition maps like those in figure \ref{Fig:igf-liquid}.  A correct treatment of the gradient energy matrices removed the asymmetry.}  For an $N$-phase, $M$-component system, the phase gradient energy matrix $\lambda$ has $N-1$ rows and columns and the component gradient energy matrix $\kappa$ has $M-1$ rows and columns.  The gradient energy coefficients coupling the implicitly defined $N^{th}$ phase (and $M^{th}$ component) are not explicitly defined in $\lambda$ and $\kappa$, but are instead distributed across all of the coefficients.  $\lambda$ and $\kappa$ are dense versions of larger matrices, $\Lambda$ and $K$, that have a direct physical interpretation.  The complete coupling of all gradients can be written in matrix form:
\begin{equation}
 \frac{1}{2}
 \begin{bmatrix}\nabla\phi_1 & \nabla\phi_2 & \cdots & \nabla\phi_N\end{bmatrix}
 \begin{bmatrix}
  \Lambda_{11} & \Lambda_{12} & \cdots & \Lambda_{1N}\\
  \Lambda_{21} & \Lambda_{22} & \cdots & \Lambda_{2N}\\
  \vdots & \vdots & \ddots & \vdots\\
  \Lambda_{N1} & \Lambda_{N2} & \hdots & \Lambda_{NN}
 \end{bmatrix}
 \begin{bmatrix}\nabla\phi_1\\ \nabla\phi_2\\ \vdots\\ \nabla\phi_N\end{bmatrix}
 \label{Eq:Lambda}
\end{equation}
The coefficients of $\Lambda$ specify an energy penalty for every possible pair of overlapping gradients.  An analogous $M\times M$ matrix $K$ contains composition gradient energy coefficients $K_{ij}$ that penalize overlapping composition gradients.

If the phase conservation constraint $\nabla\phi_N=-(\nabla\phi_1+\nabla\phi_2+\ldots+\nabla\phi_{N-1})$ obtained from Eq. \ref{Eq:phase_constraint} is substituted into Eq. \ref{Eq:Lambda} and the matrix multiplication is performed, an expression representing the gradient energy in terms of the $N-1$ phase gradients is obtained.  The coefficients in this expression are related to the $\lambda_{\alpha\beta}$ that form the matrix $\lambda$.\footnote{The diagonal terms $\lambda_{\alpha\alpha}$ are the coefficients of the squared terms, and the off diagonal terms $\lambda_{\alpha\beta}$ are equal to the coefficients of the cross terms multiplied by $\frac{1}{2}$.}  Because of the dependence of the $N^{th}$ phase on all other phases, elimination of the $N^{th}$ row and column of $\Lambda$ distributes the gradient energy coefficients for the $N^{th}$ phase across all coefficients of $\lambda$.  Thus $\lambda$ will generally be a fully dense matrix.

The physical basis for $\Lambda$ and $K$ requires that they be symmetric positive definite matrices.  $\Lambda$ and $K$ must be positive-definite because if they had negative eigenvalues, there would be a coupling of gradients (in the direction of the corresponding eigenvector) for which an increasingly sharp interface lowers the free energy of the system, producing a physically impossible negative surface energy and rendering the evolution equations unstable.  The simplification to reach Eq. \ref{Eq:Multicomponent_simplified_expansion} and Eq. \ref{Eq:F_multiphase} also reveals that $\lambda$ and $k$ are symmetric.

\subsection{Gradient energy coefficient selection}
The free energy functions and gradient energy matrices are coupled by  Eq. \ref{Eq:F_multiphase} in a way that makes fitting gradient energies to experimental systems potentially cumbersome.  At equilibrium, the surface energy of an interface is fixed once the free energy densities and interfacial widths are specified.  Thus in principle the gradient energy coefficients could be obtained by measuring the width of both the composition and phase variations at all possible equilibrium interfaces.  For a system with $N$ phases and $M$ components, there are potentially ${N \choose 2}=\frac{1}{2}(N^2-N)$ unique phase interface widths, and ${M \choose 2}=\frac{1}{2}(M^2-M)$ unique compositional interface widths.  The number of unique widths correspond exactly to the number of upper diagonal coefficients in $\kappa$ and $\lambda$.

However, the shape of free energy functions may preclude the formation of many possible interfaces, making it impossible to determine gradient energy coefficients from equilibrium observations.  In this case some gradient energies takes on a non-equilibrium role, and it may be possible to fit the coefficients to equilibrium interface widths by assuming that some of the gradient energies are zero.  In the general case that all gradient energies in Eq. \ref{Eq:gradient_energy} are non-zero, determination of the coefficients is nontrivial.  The number of unique coefficients is significantly larger than the number of equilibrium observables.  A series of \textit{ab initio} calculations would be necessary to determine the coefficients.  For each coefficient, the increase in energy when a homogeneous system is forced to incorporate a gradient must be calculated.

\section{Component evolution}
\label{Sec:Component_evolution}
Component evolution equations are derived here for a non-ideal ternary system.  Extension to a different number of components follows the same approach but is algebraically tedious.  Parts of this derivation are drawn from work by Nauman and Balsara \cite{Nauman1989} and Nauman and He \cite{Nauman1994,Nauman2001}.

The thermodynamic condition defining equilibrium in phase-separating systems is the elimination of all chemical potential gradients.  Fickian diffusion with $\nabla c$ as the driving force applies only to the special case of an ideal system where there is no enthalpy of mixing.  Systems which undergo phase separation exhibit ``uphill diffusion'', and Fickian diffusion does not hold.

To derive component evolution equations for a system characterized by a free energy functional, it is necessary to begin with the generalized form of Fick's first law:
\begin{equation}
 \vec{J_i}=-M_i\nabla\hat{\mu}_i
 \label{Eq:Ficks_first_law}
\end{equation}
$J_i$ is the flux of component $i$, $M_i$ is its mobility, and $\hat{\mu}_i$ is its diffusion potential.  In principle $J_i$ might also depend on the diffusion potential gradients of other components besides $i$, but this is not considered here.  $M_i$ is related to the diffusivity $D_i$ by the Nernst-Einstein relation:
\begin{equation}
 M_i=\frac{D_ic_i}{RT}
 \label{Eq:Nerst-Einstein}
\end{equation}
 If $D_i$ depends weakly on composition, $c_i$ will be the leading term in the mobility expression.  It is important that mobility depend on $c_i$ for conserved quantities.  If it did not, it would be possible to have a flux of a component without any of that component being present initially.

\subsection{Generalized diffusion potential}
\label{Sec:generalized_diffusion_potential}
The free energy functional $F$ is a non-equilibrium generalization of Helmholtz free energy that includes contributions from concentration and phase gradients.  At equilibrium the functional is equal to the equilibrium free energy.  To describe kinetic evolution in a non-equilibrium system, it is necessary to define a potential that approaches the chemical potential at equilibrium.  Since $F$ is a functional, the functional derivative defines an inhomogeneous (or variational) chemical potential field that becomes uniform at equilibrium:
\begin{equation}
  \hat{\mu}_i=\left(\frac{\delta F}{\delta N_i(\vec{x})}\right)_{T,V,N_{j\ne i}}
 \label{Eq:generalized_diffusion_potential}
\end{equation}
$N_i(\vec{x})$ is the number of moles of component $i$ as a function of position.\footnote{Throughout the rest of this paper, variational derivatives will be written with the assumption that the function in the denominator depends on $\vec{x}$.}  Hat notation indicates that the inhomogeneous chemical potential $\hat{\mu}_i$ is a different quantity from the standard definition of chemical potential:
\begin{equation}
 \mu_i=\left(\frac{\partial G}{\partial N_i}\right)_{T,P,N_{j\ne i}}=\left(\frac{\partial(\min F)}{\partial N_i}\right)_{T,V,N_{j\ne i}}
\end{equation}
The inhomogeneous chemical potential is defined away from equilibrium and approaches the classical chemical potential as equilibrium is approached.  At equilibrium $F$ is minimized, $\hat{\mu}_i$ is no longer a function of position, and $\hat{\mu}_i=\mu_i$.

The fundamental relation for the ternary free energy functional $F$ at constant temperature and pressure can be written as:
\begin{equation}
 \bar{F}=\frac{F}{N}=-P\bar{V}+\hat{\mu}_1c_1+\hat{\mu}_2c_2+\hat{\mu}_3c_3
 \label{Eq:F_Euler}
\end{equation}
where $\bar{F}$ is a molar quantity, $\bar{V}$ is molar volume, and $N=N_1+N_2+N_3$ is the total number of moles in the system.  It can be shown by standard thermodynamic arguments that the $\hat{\mu}_i$ obey a generalized Gibbs-Duhem relation at constant temperature:
\begin{equation}
 \sum_i c_id\hat{\mu}_i=\bar{V}dP
 \label{Eq:Gibbs-Duhem}
\end{equation}

Application of the mole fraction constraint (Eq. \ref{Eq:mass_constraint}) to Eq. \ref{Eq:F_Euler} to eliminate $c_3$ reveals that the variational derivatives of $\bar{F}$ with respect to $c_i$ are related to differences in inhomogeneous chemical potentials:
\begin{subequations}
 \begin{equation}
  \left(\frac{\delta \bar{F}}{\delta c_1}\right)_{T,V,c_2}=\hat{\mu}_1-\hat{\mu}_3
 \end{equation}
 \begin{equation}
  \left(\frac{\delta \bar{F}}{\delta c_2}\right)_{T,V,c_1}=\hat{\mu}_2-\hat{\mu}_3
 \end{equation}
  \label{Eq:variational-derivatives}
\end{subequations}
These quantities are diffusion potentials, and may be interpreted as the energy change upon adding a small amount of $c_i$ while simultaneously removing a small amount of $c_3$.  Thus the equilibrium condition of constant inhomogeneous chemical potential is equivalent to constant diffusion potential for a system with a mass constraint.

\subsection{Evolution equations}
The derivation of evolution equations begins with the observation that when individual chemical potentials are defined, their gradients are related by the Gibbs-Duhem equation.  If local thermodynamic equilibrium is assumed, the Gibbs-Duhem equation relates gradients in chemical potential $\nabla\hat{\mu}_i$ instead of changes in chemical potential $d\hat{\mu}_i$.  Local equilibrium implies that global intensive parameters vary so slowly that small neighborhoods around a point can be considered at equilibrium.  Furthermore for solids and liquids, $\bar{V}dP$ is generally very small and can be neglected in Eq. \ref{Eq:Gibbs-Duhem} for simplicity.  The Gibbs-Duhem relation for an inhomogeneous ternary system then becomes:
\begin{equation}
 c_1\nabla\hat{\mu}_1+c_2\nabla\hat{\mu}_2+c_3\nabla\hat{\mu}_3=0
\end{equation}
The mole fraction constraint (Eq. \ref{Eq:mass_constraint}) is used to eliminate $c_3$, and the equation is rearranged to put $\nabla\hat{\mu}_1$ on the left hand side:
\begin{equation}
 \begin{split}
  \nabla\hat{\mu}_1&=(\nabla\hat{\mu}_1-\nabla\hat{\mu}_3)-c_1(\nabla\hat{\mu}_1-\nabla\hat{\mu}_3)-c_2(\nabla\hat{\mu}_2-\nabla\hat{\mu}_3)\\
  &=(1-c_1)\nabla(\hat{\mu}_1-\hat{\mu}_3)-c_2\nabla(\hat{\mu}_2-\hat{\mu}_3)
 \end{split}
\end{equation}
The variational derivatives (Eq. \ref{Eq:variational-derivatives}) can now be substituted in place of the chemical potential differences:
\begin{subequations}
 \begin{equation}
  \nabla\hat{\mu}_1=(1-c_1)\nabla\frac{\delta \bar{F}}{\delta c_1}-c_2\nabla\frac{\delta \bar{F}}{\delta c_2}
 \end{equation}
A similar procedure is used to find $\nabla\hat{\mu}_2$:
 \begin{equation}
  \nabla\hat{\mu}_2=(1-c_2)\nabla\frac{\delta \bar{F}}{\delta c_2}-c_1\nabla\frac{\delta \bar{F}}{\delta c_1}
 \end{equation}
 \label{Eq:grad_mu}
\end{subequations}

The dynamics of component diffusion is governed by a mass conservation law:
\begin{equation}
\frac{\partial c_i}{\partial t}=-\nabla\cdot \vec{J}_i
\end{equation}
Substitution of Eq. \ref{Eq:Ficks_first_law}, \ref{Eq:Nerst-Einstein}, and \ref{Eq:grad_mu} produces component diffusion equations for a ternary system:
\begin{subequations}
 \begin{equation}
  \frac{\partial c_1}{dt}=\nabla\cdot\left(\frac{D_1c_1}{RT}\left((1-c_1)\nabla\frac{\delta \bar{F}}{\delta c_1}-c_2\nabla\frac{\delta \bar{F}}{\delta c_2}\right)\right)
 \end{equation}
 \begin{equation}
  \frac{\partial c_2}{dt}=\nabla\cdot\left(\frac{D_2c_2}{RT}\left((1-c_2)\nabla\frac{\delta \bar{F}}{\delta c_2}-c_1\nabla\frac{\delta \bar{F}}{\delta c_1}\right)\right)
 \end{equation}
 \label{Eq:component_evolution}
\end{subequations}
The variational derivative $\frac{\delta \bar{F}}{\delta c_i}$ is found by applying the Euler-Lagrange equation to the free energy functional (Eq. \ref{Eq:F_multiphase}):
\begin{equation}
 n\frac{\delta \bar{F}}{\delta c_i}=\frac{\delta F}{\delta c_i}=\sum_{\alpha=1}^N\phi_\alpha\frac{\partial f_\alpha}{\partial c_i}-\sum_{j=1}^{M-1}\kappa_{ij}\nabla^2 c_j
\end{equation}

\section{Phase evolution}
Phase fractions are not coupled by thermodynamic relationships and are not conserved quantities since phases are created and destroyed during phase transitions.  Thus phase evolution follows Allen-Cahn dynamics \cite{Allen1979}:
\begin{equation}
 \frac{\partial \phi_\alpha}{\partial t}=-r_\alpha\frac{\delta \bar{F}}{\delta \phi_\alpha}
 \label{Eq:phase_evolution}
\end{equation}
where $r_\alpha$ is a kinetic coefficient associated with how quickly the $\alpha$-phase can transform to another phase at constant composition.  $\frac{\delta \bar{F}}{\delta \phi_\alpha}$ is found by applying the Euler-Lagrange equation to the multiphase multicomponent free energy functional (Eq. \ref{Eq:F_multiphase}):
\begin{equation}
 \frac{\delta \bar{F}}{\delta\phi_\alpha}=\frac{\partial f_0}{\partial\phi_\alpha}-\sum_{\beta=1}^{N-1}\lambda_{\alpha\beta}\nabla^2\phi_\beta
\end{equation}
with $f_0$ defined in Eq. \ref{Eq:f_barrier}.  The implicitly defined phase fraction $\phi_N$ is a function of the other phase fractions such that $\frac{\partial\phi_N}{\partial \phi_\alpha}=-1$.  Thus the driving force for phase separation becomes $f_\alpha-f_N$, where $f_N$ is the free energy density of the implicitly defined $N^{th}$ phase.

\subsection{A barrier function for phase fractions}
\label{Sec:phase_barrier_function}
The definition of the phase fraction as a positive quantity less than or equal to 1 imposes a constraint on the phase evolution equations which was not included in their derivation.  In fact, negative phase fractions would be energetically favorable if they had physical meaning.\footnote{In systems where borrowing is allowed, negative percentages have meaning.  Consider financial leveraging - taking out a loan to make an investment.  It could be profitable to say, invest 150\% of your income by taking out a -50\% loan, if you expect the return on the investment to be higher than the interest due on the loan.}  Consider a single component system with a high energy phase $\alpha$ and a low energy phase $\beta$.  Converting $\alpha$ to $\beta$ decreases free energy by $f_\beta-f_\alpha$.  If negative phase fractions are not prohibited, there is an arbitrage where simultaneously producing more $\beta$ and negative $\alpha$ lowers the free energy without violating the phase fraction constraint.  The problem is that the global energy minima are unbounded in $\phi$.  Phase-field models typically address this issue by constructing $f(c,\phi)$ so that it has minima at $\phi=0$ and $\phi=1$ and  penalizes $\phi<0$ and $\phi>1$.  However, constructing such an interpolation for an arbitrary number of phases is problematic.

Barrier methods are often applied to minimization problems subject to inequality constraints.  Constrained optimization consists of minimizing the original potential plus the barrier functions representing the inequality constraints.  Logarithms\footnote{The $c\ln(c)$ terms in the ideal entropy of mixing are a barrier function for components that has a thermodynamic justification.} and $\frac{1}{x}$ functions are commonly used barriers, but are not ideal candidates for phase fractions which spend a lot of time in the vicinity of $\phi=0$ where the barriers are undefined.  Single phase-regions in a multiphase system would be unstable for instance, as would any evolution directed along the boundary of the feasible region, corresponding to a phase transition.

A multi-obstacle barrier \cite{Blowey1993,Garcke1999,Nestler2005} is used here to constrain phase fractions.  The barrier is zero for permissible phase fractions, and infinite otherwise.  The multi-obstacle barrier is a generalization of the double obstacle barrier used in Access models.  The double-obstacle potential was studied by Blowey and Elliott \cite{Blowey1993} and found to be consistent with curvature dependent phase boundary motion in two-phase systems.  An algorithmic implementation of the barrier is presented here for a system of $N$ order parameters that obey a constraint like Eq. \ref{Eq:phase_constraint}.

The phases in an $N$-phase system form the vertices of an $N$-simplex, and the feasible set of phase fractions lie on or within this simplex.  Enforcing that all $N$ phase fractions remain positive is enough to insure that all $N$ phase fractions will also be less than 1 because of the phase fraction constraint (Eq. \ref{Eq:phase_constraint}).  The multi-obstacle barrier is implemented by projecting a vector of phase fractions back onto the surface of the simplex when one or more phase fractions become negative as a result of advancing the evolution equations.  For a 2-phase system there is only one independent phase fraction, and the projection is trivial.  If $\phi<0$ set $\phi=0$, but if $\phi>1$ set $\phi=1$.

\begin{figure}[t]
 \includegraphics[width=.75\columnwidth]{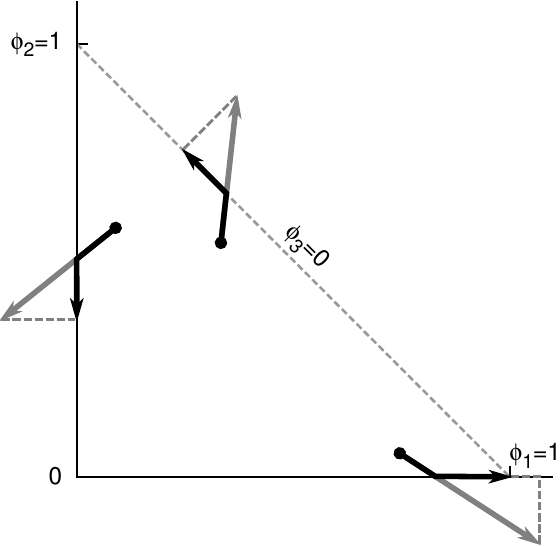}
 \caption{The multi-obstacle projection for a three-phase system.  Black points indicate initial locations in phase-space, and gray arrows a possible trajectory in the absence of constraints.  Constrained evolution proceeds along the bent black arrows.}
 \label{Fig:phase_constraint}
\end{figure}

Figure \ref{Fig:phase_constraint} offers a geometric description of the projection for a three-phase system.  Orthogonal axes are drawn to represent the two independent phase fractions $\phi_1$ and $\phi_2$, and each coordinate in the graph corresponds to a unique point in phase space.  $\phi_3=1$ at the origin, and $\phi_3=0$ corresponds to the dashed line connecting $\phi_1=1$ and $\phi_2=1$.  The constraint that all three $\phi$ be positive restricts the feasible region to the triangle with vertices at the origin, $\phi_1=1$, and $\phi_2=1$.  Any point outside of this triangle is non-physical and is given an infinite energy penalty by setting the offending phase fractions to zero.  In the case that $\phi_3$ becomes negative, the projection is accomplished by moving in the $(-1,-1)$ direction until $\phi_3$ becomes zero.

Generalization of the projection procedure for an N-phase system involves fixing violations and then recursively projecting the system to lower dimensions to fix additional violations.  The implementation of this recursive procedure is presented in algorithm \ref{Alg:multiObstacle}.

\begin{algorithm}[H]
 \caption{multiObstacle($\{\phi_1,\ldots\phi_{N-1}\}$)}
 \label{Alg:multiObstacle}
 \begin{algorithmic}
  \FOR{$\phi_i=\phi_1\ldots\phi_{N-1}$}
   \IF{$\phi_i<0$}
    \STATE $\phi_i=0$
   \ENDIF
  \ENDFOR
  \STATE $\phi_N\leftarrow 1-\sum_{i=1}^{N-1}\phi_i$
  \IF{$\phi_N<0$}
   \FOR{$\phi_i=\phi_1\ldots\phi_{N-2}$}
    \STATE $\phi_i\leftarrow\phi_i+\frac{\phi_N}{N-1}$
   \ENDFOR
  \STATE multiObstacle($\{\phi_1,\ldots\phi_{N-2}\}$)
  \STATE $\phi_{N-1}\leftarrow 1-\sum_{i=1}^{N-2}\phi_i$
  \ENDIF
 \end{algorithmic}
\end{algorithm}

\section{Results}
Experiments have repeatedly shown that liquids can often be supercooled before they solidify, but solids can almost never be superheated \cite{Dash1999,Mei2007,Rettenmayr2009}.  Solids often begin to melt below the bulk melting temperature, with liquid appearing first at triple junctions and then at grain boundaries \cite{Hsieh1989}.  Explanations have included the observation that grain boundaries and triple junctions are high energy sites that are less thermally stable than the bulk \cite{Raj1990}, that free surfaces may premelt due to atomic thermal vibrations \cite{Dash1999}, and that premelting my result from a structural transition \cite{Tang2006}.  The results in this section demonstrate premelting in nanostructures due to the shape and position of metastable free energy surfaces, possibly explaining why some experimental systems form stable liquid films at phase boundaries while others do not.

\begin{figure}[t]
 \centerline{
  \subfloat[Below the melting point, $\Delta\tilde{f}_m=1.45$.]{
    \includegraphics[width=0.55\columnwidth]{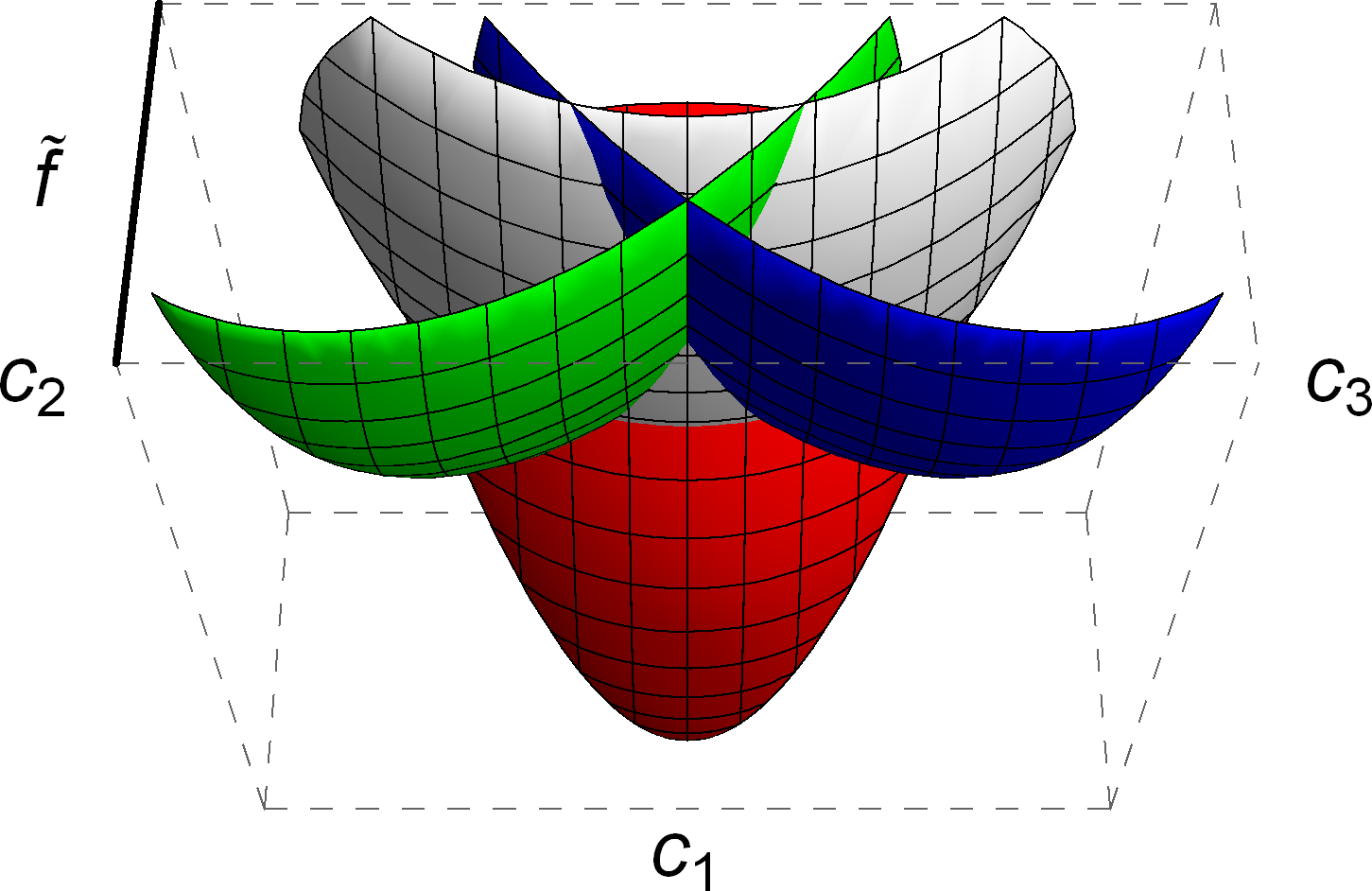}
    \includegraphics[width=0.45\columnwidth]{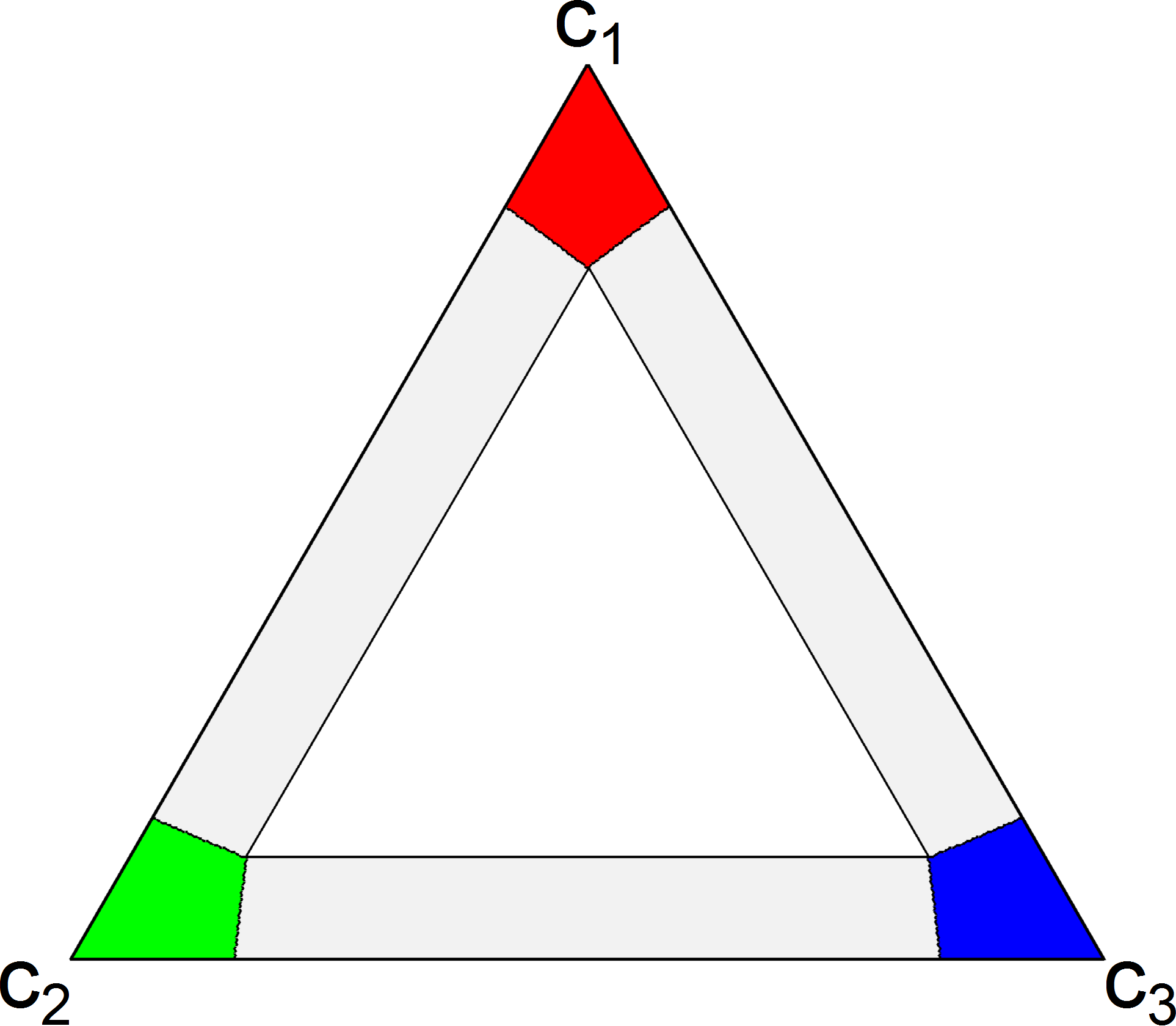}
    \label{Fig:low_T}}}
  \centerline{
  \subfloat[Above the melting point, $\Delta\tilde{f}_m=-.2$.]{
    \includegraphics[width=0.55\columnwidth]{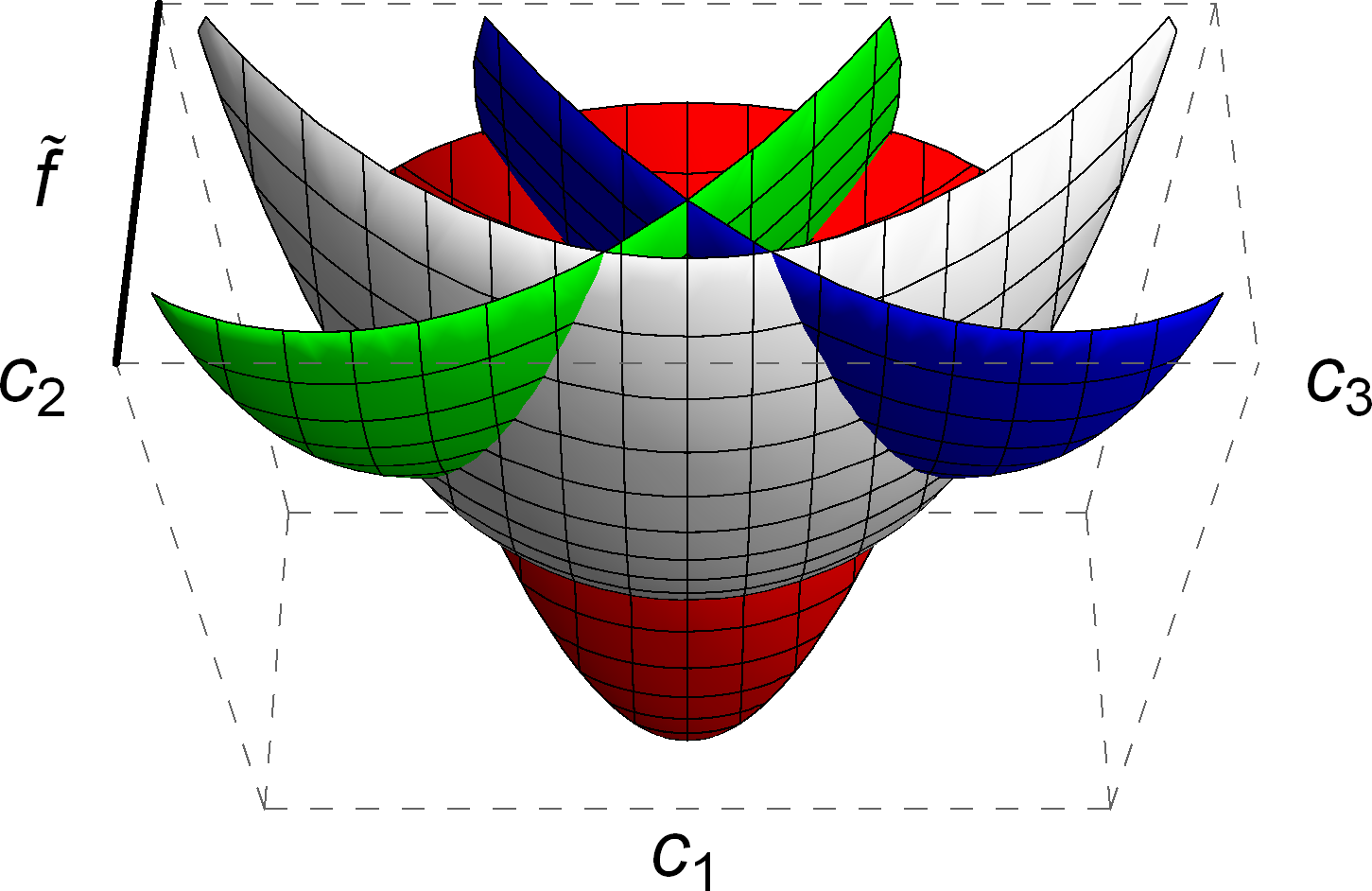}
    \includegraphics[width=0.45\columnwidth]{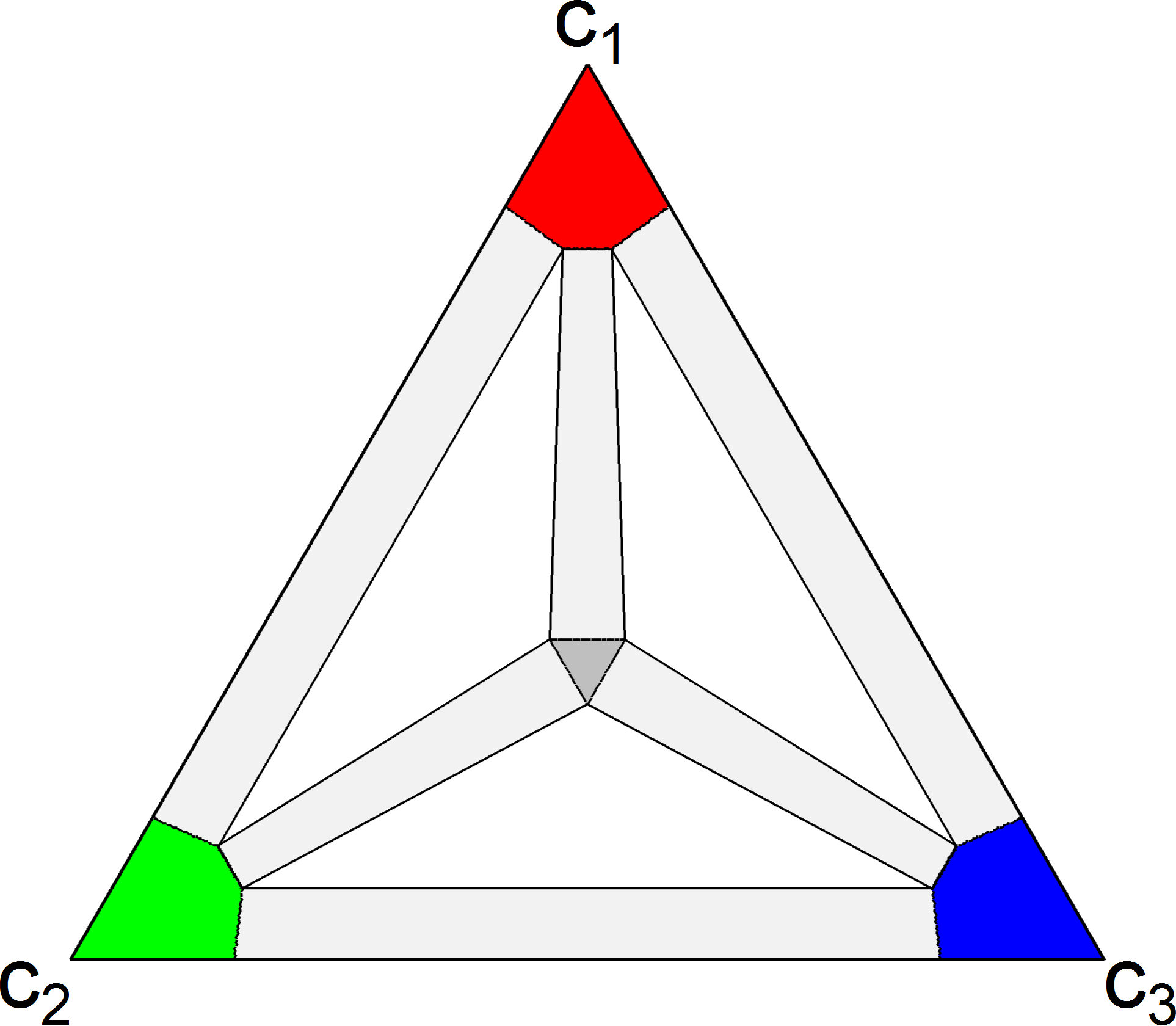}
    \label{Fig:high_T_free_energy}}}
 \caption{(Color online) Ternary eutectic free energy surfaces, viewed from below, and the corresponding phase diagrams.  The diffusing components are $c_1$ (red), $c_2$ (green) and $c_3$ (blue), and there are three solid phases: a red-rich phase, a green-rich phase, and a blue-rich phase.  A silver liquid phase appears in the center of the phase diagram when $\Delta\tilde{f}_m<0$ .}
 \label{Fig:free_energy_landscape}
\end{figure}

A four-phase ternary eutectic free energy landscape was developed from a ternary regular solution model of the form:
\begin{equation}
 \begin{split}
  \bar{f}(c_1,c_2)=&\Omega_{12}c_1c_2+\Omega_{13}c_1c_3+\Omega_{23}c_2c_3\\
  +&RT\left(c_1\ln(c_1)+c_2\ln(c_2)+c_3\ln(c_3)\right)
 \end{split}
 \label{Eq:bowl}
\end{equation}
where $ c_3=1-c_1-c_2$, and $\Omega$ determines the enthalpic contribution to free energy.  The function has a minimum at $(c_1,c_2,c_3)=(\frac{1}{3},\frac{1}{3},\frac{1}{3})$, and phases with different equilibrium compositions were created by translating Eq. \ref{Eq:bowl}:
\begin{equation}
 \begin{split}
  \bar{f}_1(c_1,c_2)&= \bar{f}(c_1+1/3-.9, c_2+1/3-.05)\\
  \bar{f}_2(c_1,c_2)&= \bar{f}(c_1+1/3-.05, c_2+1/3-.9)\\
  \bar{f}_3(c_1,c_2)&= \bar{f}(c_1+1/3-.05, c_2+1/3-.05)\\
  \bar{f}_4(c_1,c_2,T)&= \bar{f}(c_1,c_2)+.5266+\Delta\bar{f}_m\\
 \end{split}
\end{equation}
These four surfaces are plotted in figure \ref{Fig:free_energy_landscape}.  The system is a ternary eutectic in the sense that a silver liquid phase appears in the center of the phase diagram above the melting point, and upon cooling, separates into three solid phases, each with a limited amount of solubility.  The energy of the liquid surface minimum relative to the other surfaces is specified by $\Delta \bar{f}_m$, the free energy change upon melting of the solid.\footnote{If there are assumed to be no compositional effects that contribute to asymmetry in latent heat $\Delta H_m$, the change in free energy upon solidification is related to undercooling for small $\Delta T=T_m-T$ according to:
\begin{equation}
 \Delta f_m\approx\frac{\Delta H_m\Delta T}{T_m}
 \label{Eq:dG}
\end{equation}}  The liquid surface is calibrated so that the minima of all four surfaces lie on a common tangent when $\Delta \bar{f}_m=0$, corresponding to $T=T_m$.

Although the liquid surface in \ref{Fig:low_T} lies above the convex hull of the solid surfaces, there is a small region of composition space at  the center of the ternary triangle where the liquid surface is lower in energy than any of the solid surfaces at the same composition.  Liquid in this region of composition is metastable with respect to phase-separation into the three solid phases.  The simulations that follow examine the effect of such a metastable region on microstructure evolution.

The evolution equations (Eq. \ref{Eq:component_evolution} and \ref{Eq:phase_evolution}) were nondimensionalized as follows:
\begin{align*}
 \tilde{D_i}&=D_i\frac{\tau}{L^2} & \tilde{\kappa}_{ij}&=\kappa_{ij}\frac{\bar{V}}{RT L^2} &  \tilde{f}&=f\frac{\bar{V}}{RT}=\frac{\bar{f}}{RT}\\
 \tilde{r}_\alpha&=r_\alpha\frac{\tau RT}{\bar{V}} & \tilde{\lambda}_{\alpha\beta}&=\lambda_{\alpha\beta}\frac{\bar{V}}{RT L^2} &
\end{align*}
where $L$ is the characteristic length scale,  $\tau$ is the characteristic time scale for diffusion in the liquid,  $RT$ is the characteristic energy scale, and $\bar{V}$ is molar volume.  In this work  $K_{ij}$ and $\Lambda_{\alpha\beta}$ were taken to be diagonal and constant, and the following  parameters were used: $\tilde{D}_i=16$, $\tilde{r}_{\alpha}=1$, $\tilde{\Omega}_{12}=\tilde{\Omega}_{13}=\tilde{\Omega}_{23}=-10$, $\tilde{W}_{\alpha\beta}=.2$, $\tilde{K}=\tilde{\Lambda}=8$.  A large negative $\tilde{\Omega}$ insures that phases have limited solubility and there is a large energy barrier between phases in composition-space.  Choosing dimensional units of $L=\unit[1]{nm}$, $\bar{V}=\unit[10]{cm^3}$, $\tau=1.6\times\unit[10^{-9}]{s}$, $n=\unit[1]{mol}$, and $RT=\unit[8.314]{kJ/mol}$, the diffusivity in the liquid is $D=\unit[10^{-4}]{cm^2/s}$, and the equilibrium interfacial width is about \unit[8]{nm} in $c$ and \unit[4]{nm} in $\phi$.  The surface energy between solid phases is $\sigma_{ss}=\unit[1.3]{J/m^2}$, and the solid-liquid surface energy is $\sigma_{sl}=\unit[.7]{J/m^2}$ at $\Delta\tilde{f}_m=0$.

All simulations were performed on a computational grid of $512\times 512$ points using a time-adaptive pseudo-spectral method  \cite{Cogswell2010} that included Langevin noise.  Because the simulations are 2D, the system is effectively a thin film.

\subsection{Nucleation and growth}
\label{Sec:nucleation_and_growth}

\begin{figure}[t]
 \centerline{
  \subfloat[$\tilde{t}=15$]{\includegraphics[width=.45\columnwidth]{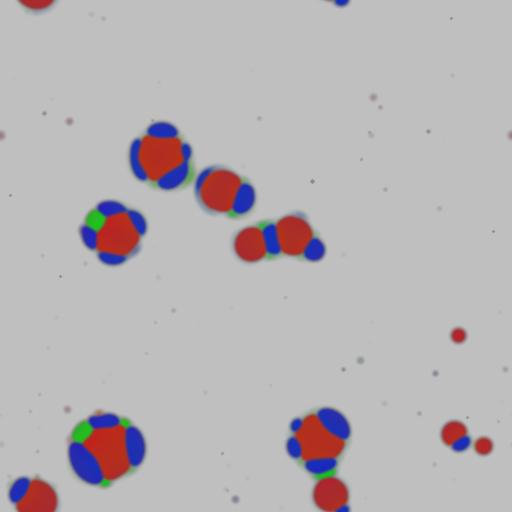}}
  \hspace{.01\textwidth}
  \subfloat[$\tilde{t}=30$]{\includegraphics[width=.45\columnwidth]{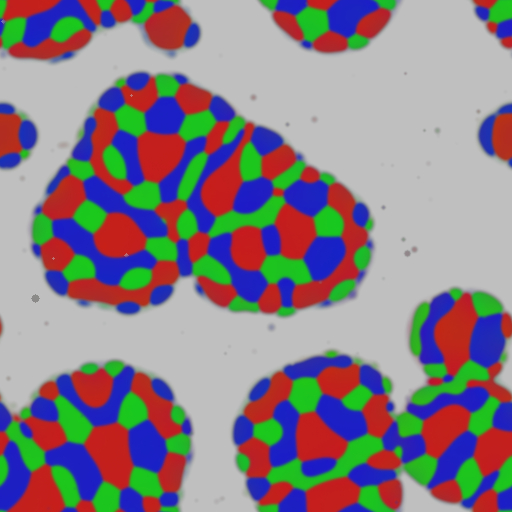}}}
 \centerline{
  \subfloat[$\tilde{t}=45$]{\includegraphics[width=.45\columnwidth]{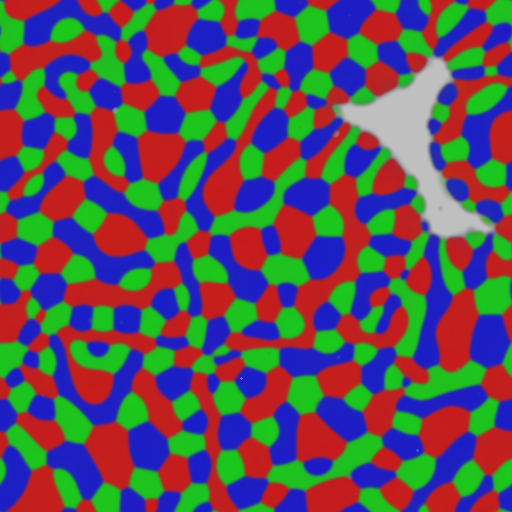}}
  \hspace{.01\textwidth}
  \subfloat[$\tilde{t}=100$]{\includegraphics[width=.45\columnwidth]{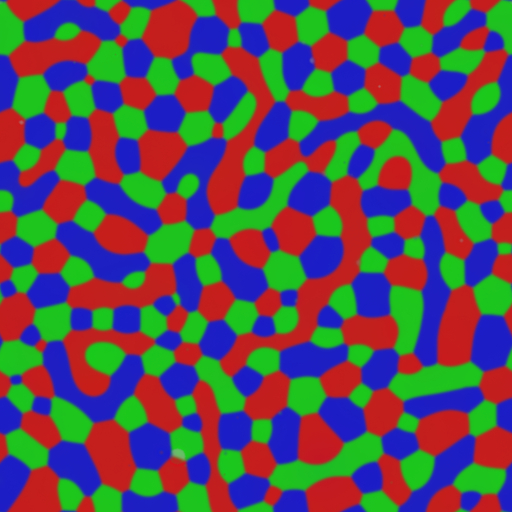}}}
 \caption{(Color online) Simulation of nucleation, growth, and coarsening of a three-phase solid from a homogeneous metastable liquid.}
 \label{Fig:nucleation_and_growth}
\end{figure}

Nucleation and growth in a ternary eutectic was simulated in figure \ref{Fig:nucleation_and_growth}.  The undercooling of the system corresponds to $\Delta\tilde{f}_m=1.45$ (figure \ref{Fig:low_T}).  The initial condition was homogeneous metastable liquid of composition $c=(.35,.31,.34)$, and all three solid phases exist at approximately equal volume fractions at equilibrium.

Langevin noise was added to the composition variables, and circular seed nuclei were added to the phase variables.  The energy of these nuclei followed a Gaussian distribution, and the radii was estimated using a classical nucleation approach as described in  \cite{Cogswell2010}.  The energy distribution and frequency of these nuclei was chosen so that phase transformation occurs in a reasonable amount of simulation time, and therefore is not rigorous.  Given the size of the system being simulated, the nucleation rate is quite large, corresponding to a system with a high density of heterogeneous nucleation sites.

Since the system is slightly enriched in $c_1$ (red component), the red-rich phase has the lowest nucleation barrier and is observed to nucleate first, as predicted by the rule of Stranski and Totomanow.  The growing nuclei then undergo secondary nucleation at the growth front and blue and green solid phases are observed to form.  Eutectic colony morphologies are expected in dilute ternary systems \cite{Plapp2002}, but there virtually no theory for multiphase morphology in concentrated systems, as noted in \cite{Hecht2004,Asta2009}.  Figure \ref{Fig:nucleation_and_growth} indicates that complex pattern formation is possible when three-phase solidification is confined to a 2D film and all phases occur at approximately equal volume fractions at equilibrium.

\subsection{Premelting and metastable liquid}

\begin{figure}[t]
 \centerline{
  \subfloat[Just before the temperature increase.]{
   \includegraphics[width=\columnwidth]{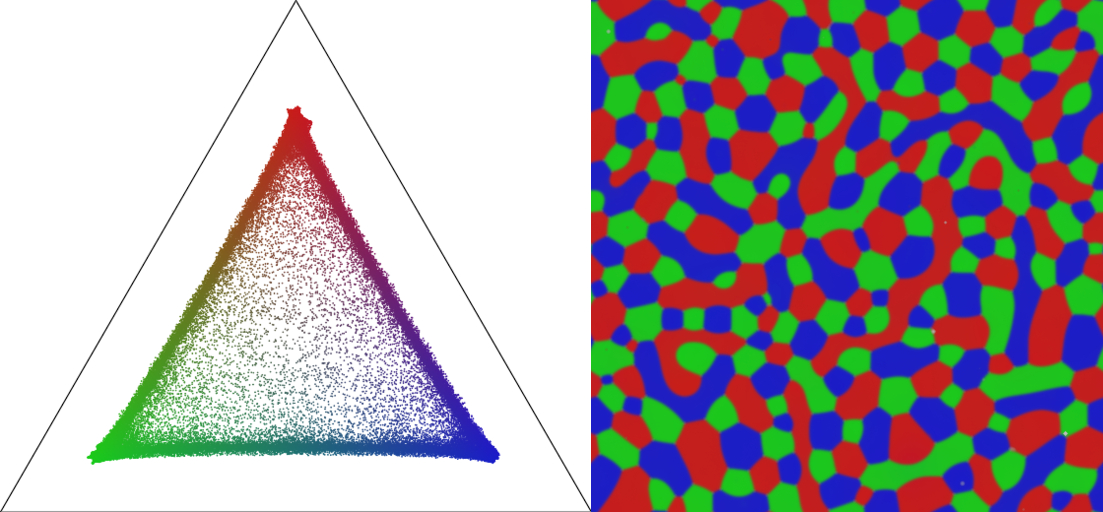}
    \label{Fig:coarsen}}}
 \centerline{
  \subfloat[Shortly after the temperature increase, $\tilde{t}=10$.]{
   \includegraphics[width=\columnwidth]{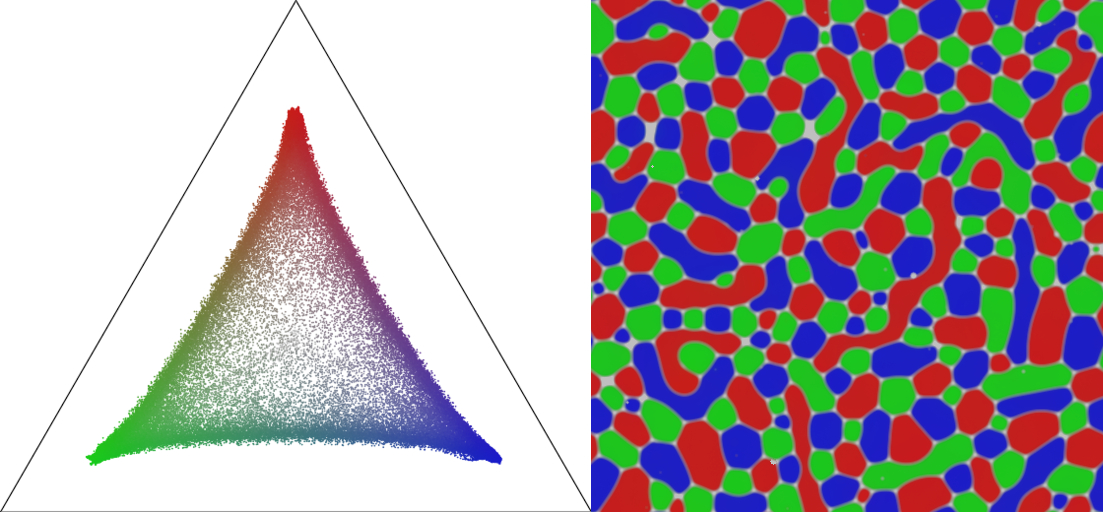}
   \label{Fig:coarsen_igf}}}
 \centerline{
  \subfloat[The liquid phase fraction of the system at $\tilde{t}=10$.  Liquid is white and solid is black.]{
   \includegraphics[width=.5\columnwidth]{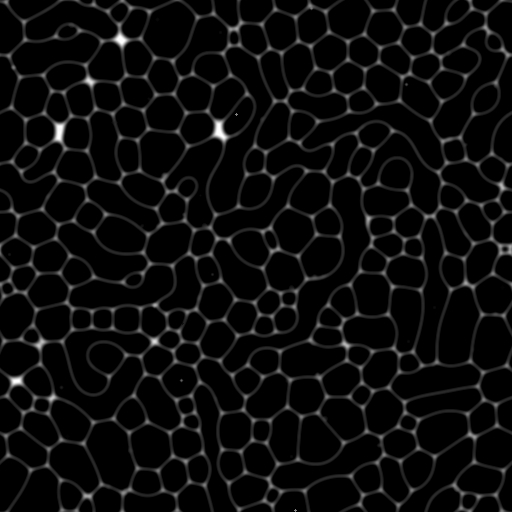}
   \label{Fig:igf-liquid}}}
 \caption{(Color online) Premelting is observed at triple junctions and phase boundaries when a three-phase solid is heated to $\Delta\tilde{f}_m=.3$, slightly below the melting point.}
 \label{Fig:igf}
\end{figure}

The completely solidified structure in figure \ref{Fig:nucleation_and_growth} was then brought to a higher temperature that was still below the melting point, and allowed to coarsen.  Because isothermal conditions are assumed, the temperature increase happens instantaneously.  Figure \ref{Fig:igf} shows the system just before and just after the temperature increase.  The magnitude of the temperature increase was not large enough to produce a stable liquid region in the phase diagram, but liquid is observed to form anyway, pooling first at phase triple junctions and then wetting phase boundaries as temperature is increased.  This behavior is qualitatively similar to experimental observations \cite{Hsieh1989, Mei2007}.

 Raj \cite{Raj1990} theorized that forming liquid at a triple junction reduces curvature and places the liquid under negative pressure, creating a stable melt pocket.  Here we find that the shape and positions of the free energy surfaces also plays an important role as well.  The liquid surface lies above the convex hull of the solid curves, but below the solid surfaces themselves over a large composition range.  Premelting may be understood as the system making use of these metastable liquid states, first at triple junctions where the energy difference between the solid and metastable liquid surfaces is largest, and then at grain boundaries where the difference is smaller but still favors liquid over compositions far from the single phase solid regions.

The composition maps\footnote{Phase-field simulations contain a lot of important information that must be extracted from images of microstructure.  To address this difficulty, a composition map was developed to visually reveal information about compositions in a ternary microstructure.  The composition map is a triangle drawn to correspond to the phase diagram, with $c_1$ (red) at the top vertex, $c_2$ (green) at the lower left, and $c_3$ (blue) at the lower right.  For every composition in the microstructure, a corresponding point is drawn on the composition map.  The color of each point matches the color of that composition in the microstructure.  For clarity, compositions that are in the liquid phase are colored silver.} in figure \ref{Fig:igf} reveal the effect of the metastable liquid surface when temperature is increased.  In figure \ref{Fig:coarsen}, the composition variation at the diffuse solid interfaces shows up as straight, diffuse lines that connect the single phase regions.  But when liquid forms at phase triple junctions and phase boundaries in figure \ref{Fig:coarsen_igf}, the interfacial composition profiles bow inward toward the center of the composition map.  The liquid free energy surface attracts interfacial compositions, and the trajectory of the interface through composition space changes so as to accommodate the low energy liquid states.

\begin{figure}[t]
 \centerline{
  \subfloat[Premelting is observed at phase triple junctions and phase boundaries.]{\includegraphics[width=.49\columnwidth]{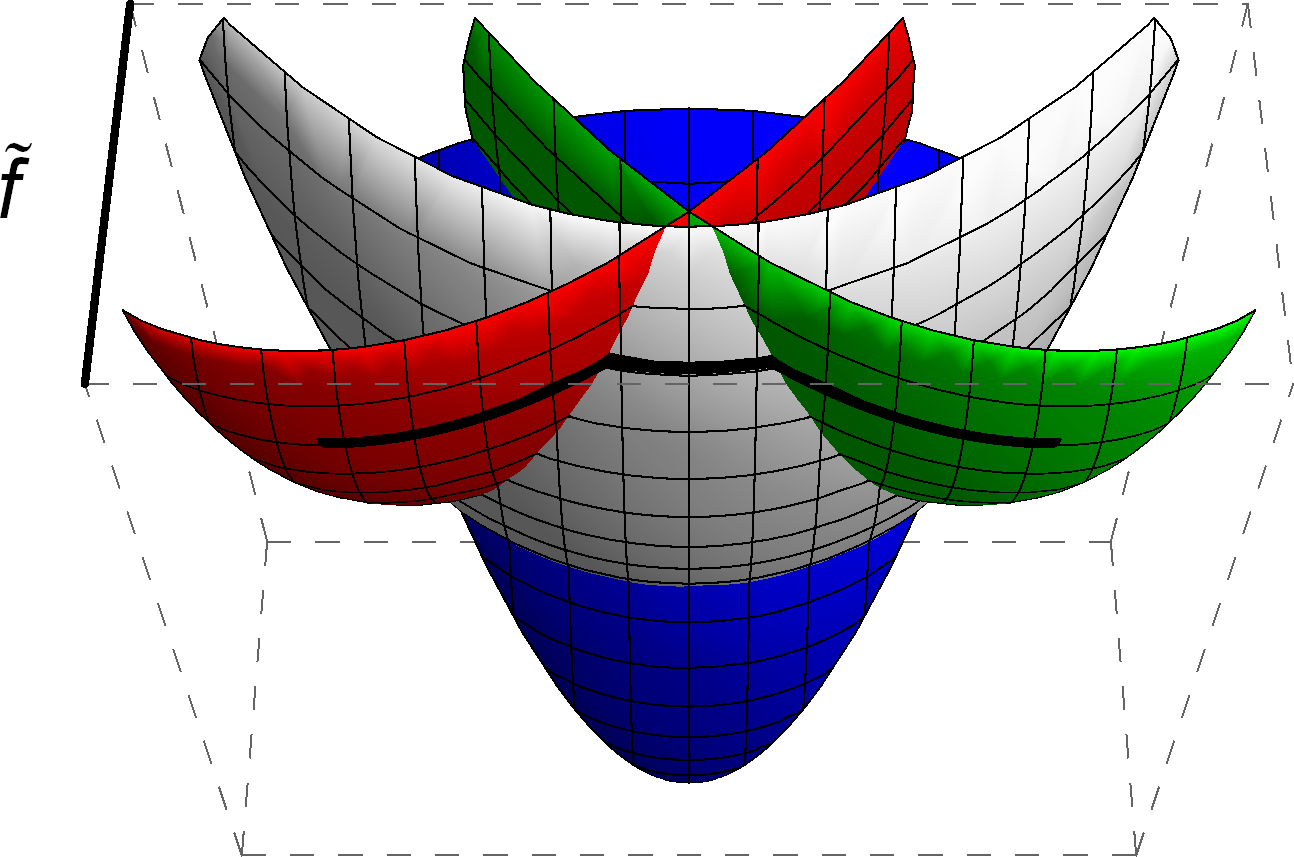}}
  \hspace{.02\columnwidth}
  \subfloat[Premelting is only observed at phase triple junctions.]{\includegraphics[width=.49\columnwidth]{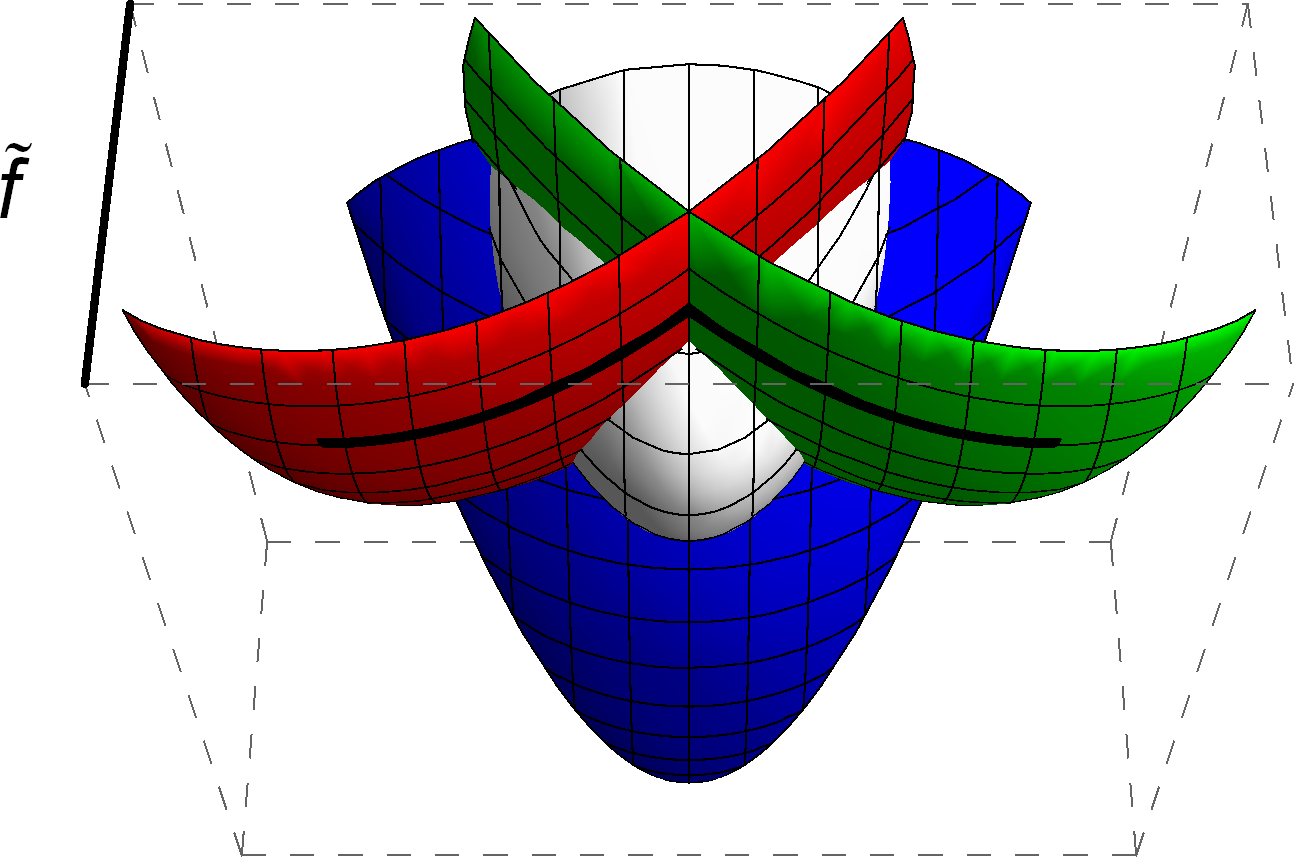}}}
 \caption{(Color online) The shape of the liquid free energy surface affects whether premelting occurs at phase boundaries.  The black lines denote a low energy path connecting the bulk compositions of two phases.  Phase boundary premelting occurs when the black line traverses part of the liquid free energy surface.}
 \label{Fig:sharp_liquid_surface}
\end{figure}

The shape of the metastable regions of free energy surfaces offers a possible explanation for why premelting is not always observed experimentally, and sometimes observed at triple junctions but not phase boundaries.  When points at a diffuse interface are forced to choose between several high energy states, the shape and position of free energy surfaces become important, as illustrated in figure \ref{Fig:sharp_liquid_surface}.   When the low energy path does not traverse the liquid surface, the system either does not premelt or must pay an energy penalty to adjust its trajectory to accommodate liquid states.  Furthermore, thickening of liquid film with increasing temperature, which is observed experimentally, may be rationalized as an increasing traversal length along the liquid surface as it descends.

\subsection{Instability of small particles}
Coarsening theory predicts that the radius of shrinking particles will smoothly decrease to zero.  However, as figure \ref{Fig:igf-liquid} illustrates, shrinking particles were observed to melt when they reached a size of \unit[10-15]{nm}, approximately twice the width of the interface.  The white pockets that are apparent in figure \ref{Fig:igf-liquid} are locations where small particles melted.  These melt pockets are temporary, and are eventually consumed by the surrounding solid.

The melting of small particles is in agreement with analysis Wagner \cite{Wagner1992}, who found that at a given undercooling, there is a critical radius below which nanocrystalline materials become unstable and melt due to geometrical effects.  Applying the analysis to $\Delta\tilde{f}_m=.3$ and our simulation parameters, we calculate the diameter of a critical particle surrounded by triple junctions \cite{Raj1990} to be \unit[8.4]{nm}, which is comparable to what was observed in our simulations.  The discrepancy might be a result of the assumptions of sharp interfaces, constant surface energy, and an overly simple expression for $\Delta\bar{f}_m$ in the analysis.  A thorough understanding of the effect of diffuse interfaces on the stability of multi-junctions is left for future work.

\subsection{Asymmetry from unequal diffusivity}
Another source of asymmetry between melting and solidification is that diffusion in a liquid is usually three to four orders of magnitude faster than in a solid.  During melting the phase that forms has high diffusivity, but during solidification the phase that forms has low diffusivity.  It has been shown that the driving force for exchange of solute across the solid-liquid interface disappears when the diffusivity of the parent phase approaches zero \cite{Rettenmayr2009,Hillert2003}.  During solidification some of the driving force must be spent on trans-interface diffusion, while during melting all of the driving force goes into interface migration.  When an alloy is cooled under nonequilibrium conditions and diffusion in the solid is limited, the composition of the solid formed initially at the core of the solidifying structure is not the same as the composition at the outer edge of the structure.  Due to nonequilibrium solute distribution in rapidly solidified supersaturated solids, solutal melting below the melting point is possible.

\begin{figure*}[t]
\centerline{
 \begin{minipage}{.5\textwidth}
   \subfloat[$\tilde{t}=0$]{\includegraphics[width=.95\textwidth]{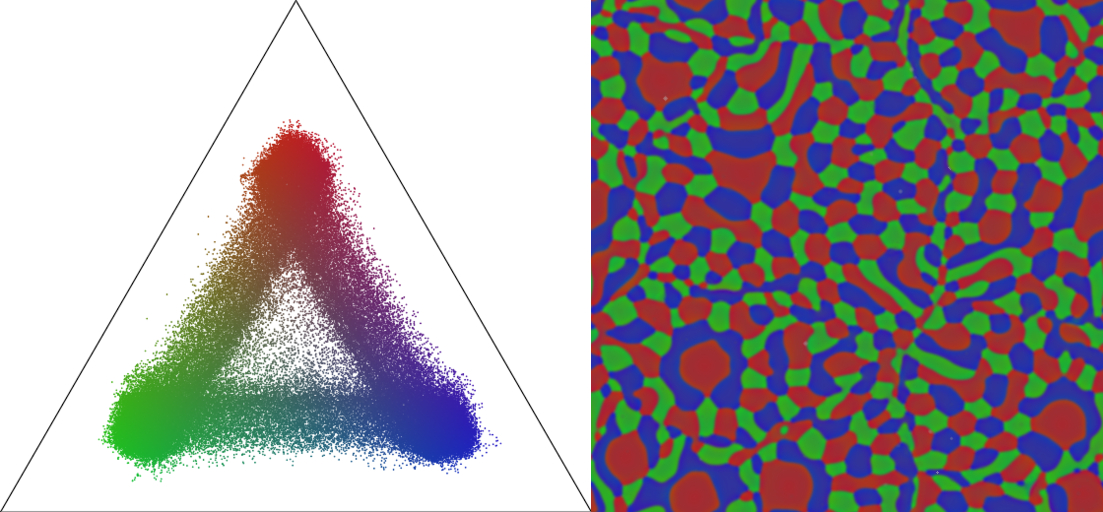}
    \label{Fig:solutal_melting_t0}}\\
   \subfloat[$\tilde{t}=5$]{\includegraphics[width=.95\textwidth]{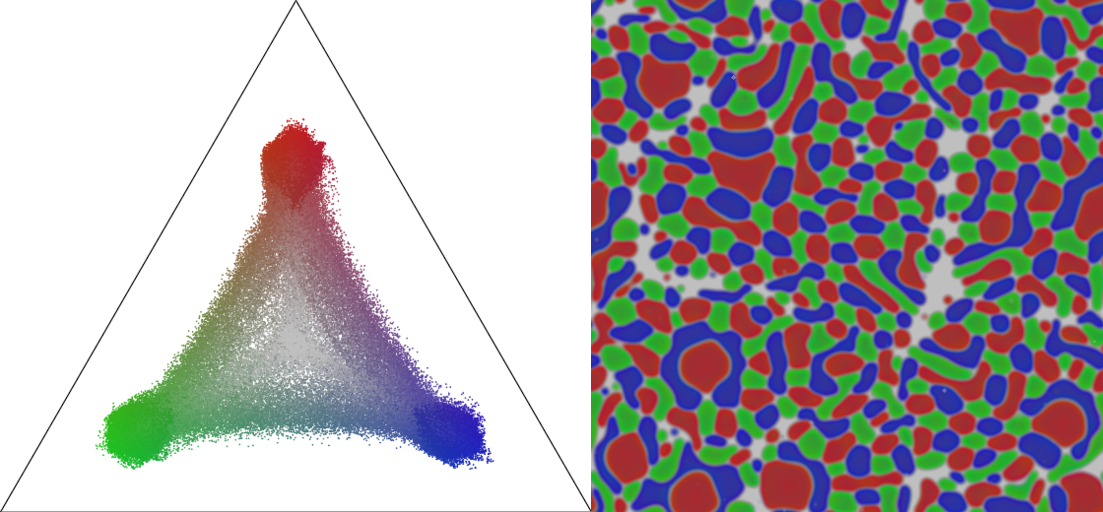}
    \label{Fig:solutal_melting_t5}}\\
   \subfloat[$\tilde{t}=50$]{\includegraphics[width=.95\textwidth]{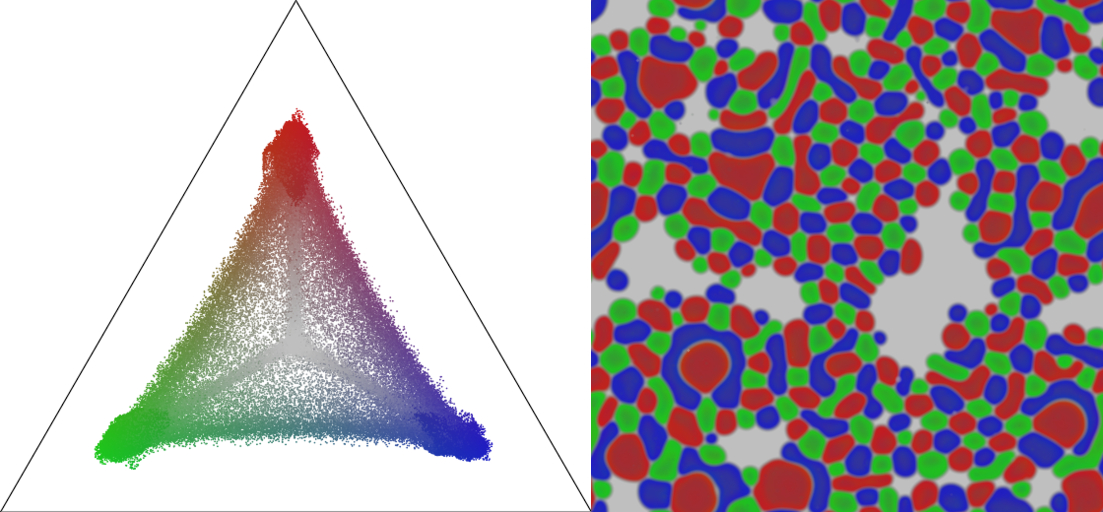}}
 \end{minipage}
 \begin{minipage}{.5\textwidth}
  \subfloat[$\tilde{t}=100$]{\includegraphics[width=.95\textwidth]{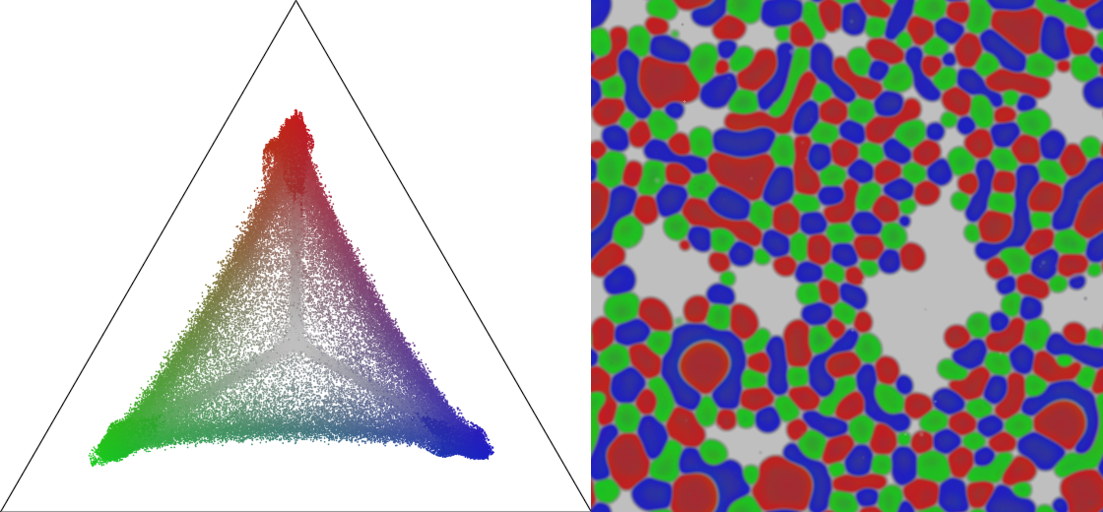}}\\
  \subfloat[$\tilde{t}=400$]{\includegraphics[width=.95\textwidth]{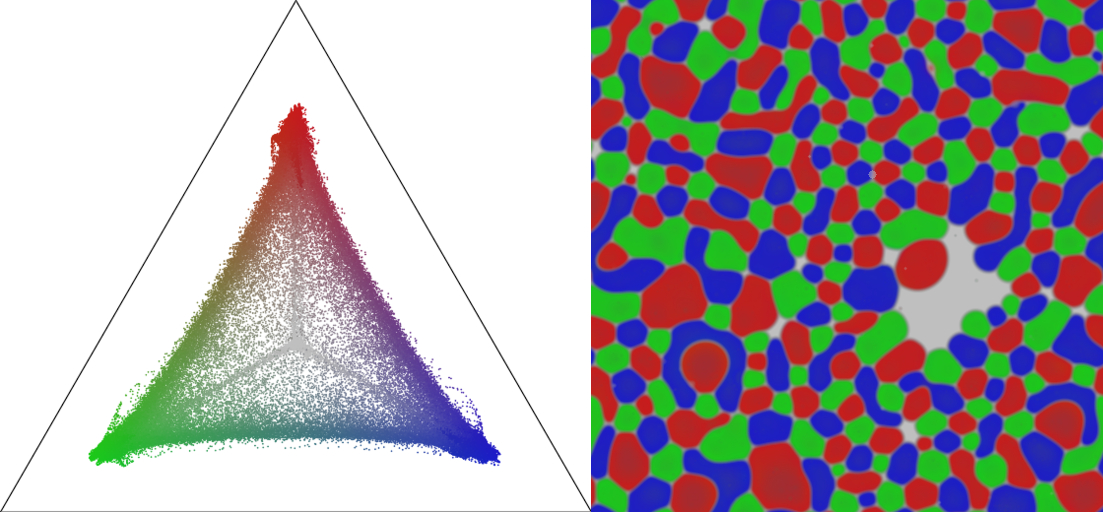}}\\
  \subfloat[$\tilde{t}=800$]{\includegraphics[width=.95\textwidth]{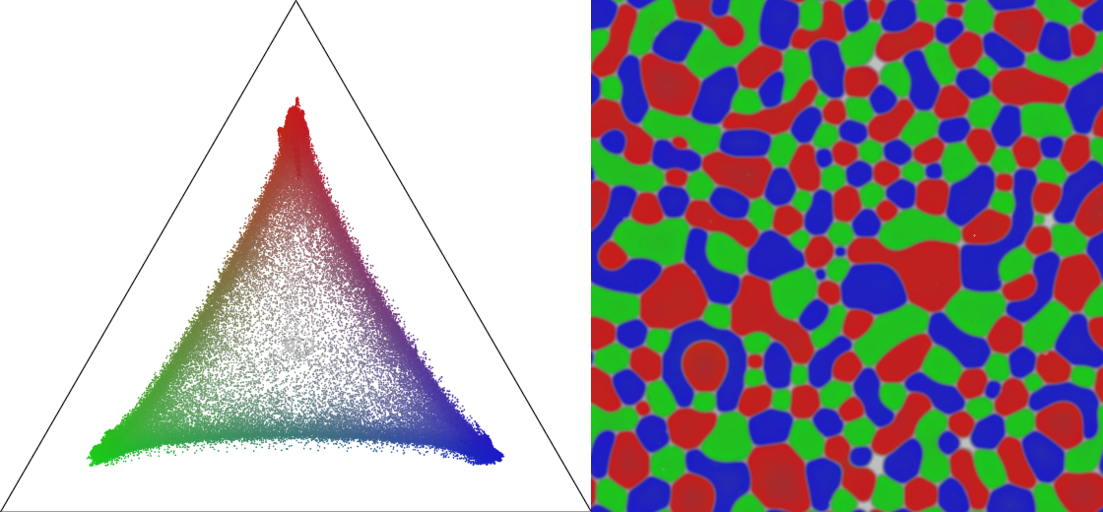}
    \label{Fig:solutal_melting_t800}}
 \end{minipage}}
 \caption{(Color online) Solutal melting and re-solidification of a rapidly solidified multiphase nanostructure held at $\Delta\tilde{f}_m=.3$, below the melting point.  The diffusivity in solid is 1/1000 the diffusivity in the liquid.}
 \label{Fig:solutal_melting}
\end{figure*}

Figure \ref{Fig:solutal_melting_t0} shows a phase-field simulation of coarsening performed with slow diffusivity in the solid.  The diffusivity of each component was a linear function of the liquid phase fraction, and diffusivity in the solid regions was decreased by three orders of magnitude.  The structure that formed consists of smaller, rougher particles that are less equiaxed.  The composition map in figure \ref{Fig:solutal_melting_t0} reveals that solidification occurred at compositions outside the stable single phase regions.  Once the system has frozen in a supersaturated state, evolution proceeds slowly because significant solid diffusion is required.

Figures \ref{Fig:solutal_melting_t5}--\subref*{Fig:solutal_melting_t800} show the system shortly after being heated to $\Delta\tilde{f}_m=.3$, which corresponds to a temperature below the melting point.  The temperature increase initially causes the regions of supersaturated solid to melt.  Large pools of liquid form, but eventually the solid surrounding these liquid regions grows back into the liquid.  When the liquid has re-solidified, the composition map appears qualitatively similar to that in figure \ref{Fig:coarsen_igf}.

\section{Conclusion}
A diffuse interface model for microstructure with an arbitrary number of phases and components was derived from basic thermodynamic and kinetic principles.  Interfaces were treated as thermodynamic entities and nonlinear diffusion equations for concentrated solutions were derived in accordance with the Gibbs-Duhem and Nernest-Einstein relations.  A composition gradient energy was included for the first time in a multiphase model to capture the effects of solute trapping, and an inhomogeneous diffusion potential was introduced as the driving force for diffusion without a dilute solution approximation.  Inhomogeneous free energy for a multicomponent, multiphase system was obtained from a Taylor expansion that produced matrices of gradient energy coefficients.  It was shown how the properties of each phase and component may be specified independently of the others, even when the phase fractions and mole fractions obey a mass constraint.  A linear interpolation between free energy surfaces was used to avoid problematic pair-wise interaction of phases, and a multi-obstacle barrier was applied to permit arbitrary barriers between phases.

The model is well-suited for studying phenomena where interfacial width is important, and captures details of melting and solidification that have not previously been modeled with phase-field methods.  A nucleation barrier to solidification was observed, and melting in solids was found to start below the melting point at phase triple junctions and phase boundaries, where pockets of liquid and stable liquid films formed.  Premelting was the result of low-energy pathways through composition space provided by metastable portions of free energy surfaces.  Small particles were observed to be unstable to heating as predicted by theory, and the large difference in diffusion constants between solid and liquid was found to lead to solutal melting, common behavior in rapidly solidified alloys.

\begin{acknowledgments}
We are grateful to Monika Backhaus-Ricoult for motivating the study of transient and metastable phases,  Samuel Allen for careful critique of the derivations, and James Warren for providing advice on barrier functions.  We also thank the anonymous reviewers of this paper for providing helpful, thoughtful, and thorough feedback.  Financial support for this research was provided by Corning Inc.
\end{acknowledgments}

\bibliography{Multiphase}
\end{document}